\documentclass[reprint,amsmath,amssymb,aps,prb,]{revtex4-2}

\usepackage{graphicx}
\usepackage{dcolumn}
\usepackage{bm}
\usepackage{braket}
\usepackage[version=4]{mhchem}
\usepackage[caption=false]{subfig}
\usepackage{epsfig}
\usepackage{ulem}
\usepackage{color}
\usepackage{mathrsfs}
\usepackage{setspace}
\usepackage{array}
\usepackage{amsmath}
\usepackage{amssymb}
\usepackage{dsfont}
\usepackage{textcomp,gensymb}
\usepackage{dcolumn}
\usepackage{multirow}
\usepackage{bibentry,natbib}
\usepackage{booktabs}

\newcommand\rmd{\mathrm{d}}

\newcommand{\hb}{\hbar}

\renewcommand\parallel{\mathrel{/\mskip-2.5mu/}}

\newcommand\pt{\partial}

\newcommand\dg{\dagger}

\newcommand{\inp}[2]{\langle #1  | #2  \rangle}

\newcommand\sg{\sigma}

\newcommand\lm{\lambda}

\begin{document}
	\title{Geometric phase driven Josephson junction: Possible experimental scheme for the search of spin superfluidity}
	
	\author{Yilin Liu$^{1,2}$}
	\author{Zi-Jian Li$^1$}
	\author{Jiadu Lin$^1$}
	\author{Qing-Dong Jiang$^{1,3}$}
	\email{qingdong.jiang@sjtu.edu.cn}
	\affiliation{{}\\ $^1$ 
		Tsung-Dao Lee Institute  \& School of Physics and Astronomy, Shanghai Jiao Tong University, Shanghai 200240, China\\
		$^2$ Department of Modern Physics, University of Science and Technology of China, Hefei, Anhui, 230026, China\\
        $^3$ Shanghai Branch, Hefei National Laboratory, Shanghai 201315, China}
  
	\begin{abstract}
	We use the Gross-Pitaevskii equation to study Josephson tunneling between two weakly coupled Bose-Einstein condensates, which compose spin-1 bosons. We show that a rotating magnetic field on one side can produce a phase difference across the junction, resulting in an oscillatory tunneling spin current. Besides numerical calculation, we derive analytical results in two extreme cases, namely the low- and high-frequency limits: in the low-frequency limit (magnetic field rotates adiabatically), a non-Abelian geometric phase arises and leads to the oscillatory spin current. By sharp contrast, the physics is intrinsically different in the high-frequency limit, where an average Zeeman energy difference leads to an oscillatory spin current. This proposed apparatus should be promising for the future experimental search of spin superfluidity.
	\end{abstract}
	\maketitle
	\section{Introduction}
	\label{Intro}
	In the past few years, spin superfluidity has attracted a lot of attention both theoretically and experimentally. Spin superfluidity (or spin superconductivity, if one would like to emphasize the response of the carrier’s spin to external electromagnetic field) is a Bose-Einstein condensation (BEC) of spin-1 bosons \cite{spin_current_review}. It may exist in many different systems, such as ferromagnetic graphene \cite{ssc_foundation_1}, BEC of magnetic atoms \cite{ssc_foundation_5,ssc_foundation_6,ssc_foundation_7,ssc_foundation_4}, \ce{$^3$He} superfluidity \cite{ssc_foundation_8,ssc_foundation_9,PhysRevB.88.094503,LEGGETT196876,PhysRevLett.129.016801}, and so on \cite{ssc_foundation_2,ssc_foundation_3,ssc_foundation_4,ssc_foundation_10,ssc_foundation_11,ssc_foundation_12,ssc_foundation_13,ssc_foundation_14,ssc_foundation_15,ssc_foundation_16,ssc_foundation_17,ssc_foundation_18,PhysRevB.99.104423,PhysRevB.95.144402}. Spin superfluidity is expected to show many exotic new phenomena such as dissipationless spin current \cite{GL-1} and the electric Meissner effect \cite{meissner}. Theoretically, a Ginzburg-Landau-type theory has been proposed for a spin-polarized \cite{GL-1} and spin-nonpolarized \cite{GL-2} superfluid, respectively. However, only in recent years has the signature of a spin-superfluid ground state been observed in real experiment \cite{ssc_expr_1,ssc_expr_2} and more experimental signature for spin superfluidity is still required \cite{spin_current_review}.
	
	On the other hand, it is well known that scientists can manipulate macroscopic numbers of bosons since the observation of the BEC of cold alkali atoms \cite{BEC_coldatom_Nobel,BEC_2}. Amongst numerous experimental techniques, this manipulation can be realized via the bosonic Josephson junction (BJJ) \cite{smerzi_BJJ_PRL}. The dynamics of spinless BJJ has been theoretically studied and many new phenomena, such as macroscopic quantum self-trapping and $\pi$-phase oscillations, have been predicted \cite{Raghavan_PRA} and experimentally verified \cite{self-trap_PRL}. For a spin-1 BEC system, people have also studied the transition between three energy states, known as the internal Josephson effect \cite{internalBJJ_1,internalBJJ_2,ssc_foundation_5}.
	
	In the zoo of miscellaneous Josephson junctions, it is well known that a current will be generated if there is a phase difference between two sides of the junction. For a junction of two superconductors weakly linked by a thin insulating layer, the current is proportional to the sine of this phase difference \cite{josephson1962,josephson_1,josephson_2}. For other Josephson junctions, the functional dependence of current on the phase difference can be more complex \cite{buzdin2005proximity,yerin2017proximity}. On the other hand, from elementary quantum mechanics, it is also well known that, in the presence of a rotating magnetic field, a spin will accumulate a geometric phase (also known as the Berry phase \cite{NQ_berryphase}). The Berry phase effects have been considered in superconducting systems \cite{geo_sc_1,geo_sc_2}. By applying a spatial magnetic field gradient, a Berry phase term can be generated in a spinor condensate system \cite{geo_int_1}. Moreover, the mesoscopic spin Berry phase effect can be realized in a coupled two-mode BEC system \cite{geo_int_1}.
	In this work, we use the Berry phase to control the particle current in the BJJ system: if we put the spin superfluid in a junction, we expect that applying a rotating magnetic field shall generate some sort of “phase difference” akin to that of the Josephson effect. Accordingly, a particle current could be generated and observed experimentally, serving as another clear signature of spin superfluidity.
	
	The BJJ offers an ideal platform for the realization of a spinor Josephson effect. Schematically, a BJJ consists of two traps which act as “containers” for two condensates \cite{self-trap_PRL}, which can be modeled by a double-well potential \cite{Raghavan_PRA} (Fig.\,\ref{SBJJ}). The physics of a BEC system can be well captured by the Gross-Pitaevskii equation (GPE) \cite{smerzi_BJJ_PRL}. We assume these two condensates are “weakly coupled,” given that the barrier of potential is high enough to suppress the tunneling matrix element.
	
	In the present paper, we extensively discuss the effect of a rotating magnetic field on a BJJ with spin-1 bosons (at zero temperature). For simplicity, at first we neglect the mutual interaction between bosons (which is weak) and the spin-orbit coupling effect (which is a relativistic correction \cite{soc_relativistic}). In that case, the GPE simply reduces to a linear, time-dependent Schr\"{o}dinger equation. Following the treatment for the spinless case \cite{Raghavan_PRA}, we integrate out the spatial dependence of the macroscopic wave function, reducing a PDE problem to a two-mode time-dependent problem. We shall then focus on two scenarios. 
	
	First, we consider the low-frequency limit, in which case the characteristic energy scale of the system’s period is much smaller than the energy spacing. So we can address the problem with the help of the adiabatic theorem. In fact, concepts from adiabatic evolution, such as the (non-Abelian) Berry connection and gauge invariance, emerge in the preceding context. It turns out that a sinusoidal oscillation of the particle number difference should be expected, which leads to a time-averaged population imbalance. 
	
	Second, we consider the high-frequency limit, in which case the energy scale of the period is larger than any other scales appearing in the system. In that case, the time-averaged evolution of the system appears to be of most importance. To extract this time-averaged information, we resort to the Floquet theory, where we obtained the effective Floquet Hamiltonian by means of the Magnus expansion \cite{Floquet_ref}. It turns out that there is also an oscillatory current (at least in a time-averaged sense) and three internal (spin) degrees of freedom are decoupled.
 
    Strictly speaking, the effect of mutual interaction among constituent bosons is non-negligible, playing a crucial role in stabilizing a superfluid phase \cite{annett_sc_book}. In Sec.\,\ref{Interaction} we substantiate the qualitative robustness of our results in the presence of the weak self-interaction among cold atoms.
	
	The paper is organized as follows. First, in Sec.\,\ref{GPE} we present the GPE of the system and show that it can be reduced to a two-mode Schr\"{o}dinger equation by applying the variational ansatz. Then, in Sec.\,\ref{LF} we discuss the low-frequency limit and calculate the particle current. In Sec.\,\ref{HF} we study the high-frequency limit via the Floquet theory and discuss the particle current in that case. Finally, the effect of self-interaction is discussed in Sec.\,\ref{Interaction}. In Appendix A we show a generalization of the (non-degenerate) adiabatic theorem. In Appendix B we show that our prediction derived in Sec.\,\ref{LF} is gauge invariant (i.e. independent of the choice of instantaneous eigenstates). Our results are summarized in Sec.\,\ref{Summ}.

	\section{The Gross-Pitaevskii Equation and the Corresponding Reduced Problem}
	\label{GPE}
	\begin{figure}[htbp]
		\centering
		\includegraphics[width=0.45\textwidth]{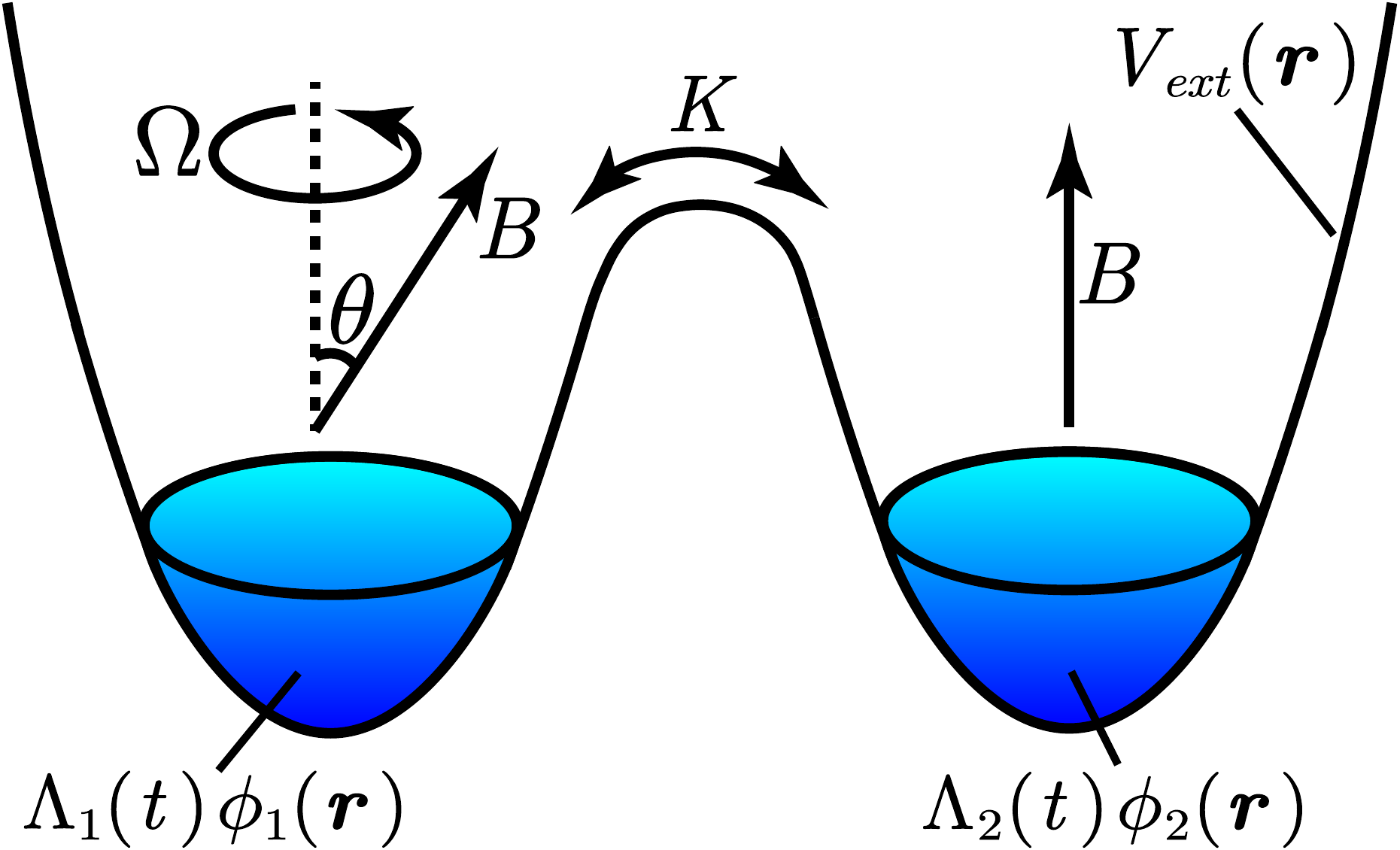}
		\caption{The apparatus for BJJ. The colored areas stand for the condensate and we use a double-well potential $V_{\mathrm{ext}}(\bm{r})$ to represent the trap. }
		\label{SBJJ}
	\end{figure}
	The apparatus of BJJ is schematically illustrated in Fig.\,\ref{SBJJ}. Since we are considering a system of spin-1 particles, the macroscopic wavefunction $\Psi(\bm{r},t)$ is a three-component spinor. The time-dependent GPE \cite{smerzi_BJJ_PRL} can be generally written as
	\begin{equation}
	i\hb\pt_t\Psi=\left(-\frac{\hb^2}{2m}\nabla^2+V_{\mathrm{ext}}(\bm{r})-g\mu_B\bm{s}\cdot\bm{B}(\bm{r},t)+F(\Psi)\right)\Psi
	\label{SBJJ_schrodinger_1}
	\end{equation}
	The first three terms of the Hamiltonian correspond to kinetic, potential, and Zeeman energy, respectively. The last term $F(\Psi)$ is some quadratic function of $\Psi$ that describes the self-interaction effect, which would be extensively discussed in the Sec.\,\ref{Interaction}. The magnetic field $\bm{B}(\bm{r},t)$ represents two localized uniform fields. The field on the left is rotating with $\theta$ the polar angle while the field on the right is static along the $z-$axis with the same magnitude. Also note that $\bm{s}$ denotes the spin operator for spin-1 bosons. As a matter of convention, we will henceforth use $\bm{s}$ to represent three Pauli matrices for spin $j=1$:
	\begin{equation}
	\begin{aligned}
	&s_x=\frac{1}{\sqrt{2}}\begin{pmatrix}
	& 1 & \\
	1&  &1\\
	& 1 &
	\end{pmatrix},\quad
	s_y=\frac{1}{\sqrt{2}}\begin{pmatrix}
	& -i & \\
	i&  &-i\\
	& i &
	\end{pmatrix},\\
	&s_z=\begin{pmatrix}
	1&  & \\
	& 0 &\\
	&  &-1
	\end{pmatrix}
	\end{aligned}
	\label{su2_lie_algebra_j=1}
	\end{equation}
	and use $\bm{\sg}$ to represent the original spin $j=1/2$ Pauli matrices.
	
	Analogous to the case of spinless BJJ, we employ the widely-known two-mode approximation \cite{two-mode_1,smerzi_BJJ_PRL,Raghavan_PRA} for $\Psi$, which is the linear superposition of two condensates' macroscopic wave functions that can be decomposed as a product of spatial and spinorial part:
	\begin{equation}
	\Psi(\bm{r},t)=\Lambda_1(t)\phi_1(\bm{r})+\Lambda_2(t)\phi_2(\bm{r})
	\label{ansatz}
	\end{equation}
	where $\phi_1$ and $\phi_2$ are (c-numbered) ground state wave functions for two separated wells (e.g., ground state wavefunction of some localized 3D harmonic potential), while $\Lambda_1$ and $\Lambda_2$ are spinors representing the internal (spinorial) degree of freedom. We assume that they are real functions (a reasonable assumption for, say, systems with harmonic potentials) normalized to unity and "almost" orthogonal. Here, by "almost orthogonal" we mean the overlap of $\phi_1$ and $\phi_2$
	\begin{equation}
	\int\rmd^3r\phi_1^*(\bm{r})\phi_2(\bm{r})=\int\rmd^3r\phi_1(\bm{r})\phi_2(\bm{r})
	\end{equation}
	is zero for most cases except when they are "sandwiching" the kinetic and potential energy:
	\begin{equation}
	\begin{aligned}
	K&\equiv-\int\rmd^3r\phi_1(\bm{r})\left(-\frac{\hb^2}{2m}\nabla^2+V_{\mathrm{ext}}(\bm{r})\right)\phi_2(\bm{r})\\
	&=-\int\rmd^3r\phi_2(\bm{r})\left(-\frac{\hb^2}{2m}\nabla^2+V_{\mathrm{ext}}(\bm{r})\right)\phi_1(\bm{r})
	\end{aligned}
	\end{equation}
	The matrix element $K$ models the tunneling effect between two well-separated potentials due to the particle's kinetic and potential energy.
	
	It is customary to normalize $\Psi$ such that its norm squared is the total particle number, instead of unity. We do not adopt this convention and continue to normalize $\Psi$ to unity, which is convenient for implementing the standard treatment of perturbation theory later on. Thus, the norm squared of $\Lambda_1$ and $\Lambda_2$ ($N_1\equiv\Lambda^{\dg}_1\Lambda_1$ and $N_2\equiv\Lambda^{\dg}_2\Lambda_2$) are ratios of particle number on the left/right to the total number, respectively.
	
	Substituting the ansatz into Eq.\,\eqref{SBJJ_schrodinger_1} and neglecting the nonlinear term, we have
	\begin{equation}
	\begin{aligned}
	&i\hb(\phi_1\pt_t\Lambda_1+\phi_2\pt_t\Lambda_2)\\
	&=\Lambda_1\left(-\frac{\hb^2}{2m}\nabla^2+V_{\mathrm{ext}}(\bm{r})\right)\phi_1\\
	&+\Lambda_2\left(-\frac{\hb^2}{2m}\nabla^2+V_{\mathrm{ext}}(\bm{r})\right)\phi_2\\
	&-\phi_1g\mu_B\bm{s}\cdot\bm{B}(\bm{r},t)\Lambda_1
	-\phi_2g\mu_B\bm{s}\cdot\bm{B}(\bm{r},t)\Lambda_2
	\end{aligned}
	\end{equation}
	Taking the overlap with $\phi_1$ and $\phi_2$, the problem now reduces to a two-mode time-dependent problem
	\begin{equation}
	\begin{cases}
	\;i\hb\pt_t\Lambda_1=-g\mu_B\bm{s}\cdot\bm{B}_1(t)\Lambda_1-K\Lambda_2&\\
	\;i\hb\pt_t\Lambda_2=-K\Lambda_1-g\mu_B\bm{s}\cdot\bm{B}_2\Lambda_2&
	\end{cases}
	\end{equation}
	where we have employed the fact that $\phi_1$ and $\phi_2$ are (approximate) ground state eigenfunctions. By assuming the double-well potential to be symmetric, one is free to set their ground state energies to zero. Also notice that the overlap integration eliminates the spatial dependence of $\bm{B}(\bm{r},t)$, reducing it into two localized fields, $\bm{B}_1(t)$ and $\bm{B}_2$.
	Note that $\bm{B}_1(t)=B\bm{n}(t)$ and $\bm{B}_2=B\bm{e}_z$ 
	[with $\bm{n}(t)=(\sin\theta\cos\Omega t,\sin\theta\sin\Omega t,\cos\theta)^T$ a rotating direction vector] and we can align $\Lambda_1$ and $\Lambda_2$ to form a six-component object; therefore, the problem now becomes solving the following time-dependent problem with a $6\times 6$ Hamiltonian:
	\begin{equation}
	i\hb\pt_t\begin{pmatrix}
	\Lambda_1\\
	\Lambda_2
	\end{pmatrix}
	=
	\begin{pmatrix}
	-g\mu_BB\bm{s}\cdot\bm{n}(t) & -K\\
	-K & -g\mu_BBs_z
	\end{pmatrix}
	\begin{pmatrix}
	\Lambda_1\\
	\Lambda_2
	\end{pmatrix}
	\end{equation} 
	
	For the convenience of doing numerical calculations, we adopt the dimensionless time ($\tau$), tunneling energy ($k$) and angular velocity ($\omega$), defined as
	\begin{equation}
	\tau\equiv\frac{g\mu_B B}{\hb}t,\quad k\equiv\frac{K}{g\mu_B B},\quad \omega\equiv\frac{\hb}{g\mu_BB}\Omega.
	\end{equation}
  However, we will come back to using dimensional parameters when discussing physical implications of our predictions.
	Thus the time-dependent problem is now expressed as
	\begin{equation}
	i\pt_{\tau}\begin{pmatrix}
	\Lambda_1\\
	\Lambda_2
	\end{pmatrix}
	=
	\begin{pmatrix}
	-\bm{s}\cdot\bm{n}(\tau) & -k\\
	-k & -s_z
	\end{pmatrix}
	\begin{pmatrix}
	\Lambda_1\\
	\Lambda_2
	\end{pmatrix}
	\label{SBJJ_reduced}
	\end{equation}
	with $\bm{n}(\tau)=(\sin\theta\cos\phi,\sin\theta\sin\phi,\cos\theta)^T$ ($\phi\equiv\omega\tau$ is the azimuthal angle). Note that the Hilbert space is six dimensional, because it combines three-dimensional spinors on each side of the BEC.

	\section{The Low-Frequency Limit}
	\label{LF}
	
	\subsection{Adiabatic Dynamics}
	Let us first consider the case of the low-frequency limit where the frequency $\omega$ is sufficiently low. Given that the system is weakly coupled, one can treat the tunneling term $k$ as perturbation. Without this perturbation, the eigenvalues of Hamiltonian are $\pm 1$ and $0$. When the perturbation is applied, one should expect the removal of degeneracy and energy split of order $\sim O(k)$. To summarize, this is a periodic system with low frequency and an ``almost" degenerate spectrum. This reminds us of the concept of adiabatic evolution.
	
	First, we consider a generic time-dependent system with Hamiltonian $\mathcal{H}=\mathcal{H}(\lm(\tau))$. Here, $\lm=\lm(\tau)$ means that the time dependence of the Hamiltonian is realized through time-dependent parameters defined in parameter space $\{\lm^i|i=1,\cdots, N\}$ with $N$ the total number of parameters. Instantaneous eigenstates are denoted as $\ket{n_i(\lm=\lm(\tau))}$. Just like the system we are interested in, we assume that the Hamiltonian can be written as a sum of unperturbed and perturbation parts and it has an "almost" degenerate spectrum (in the sense that the possibly time-dependent energy splits are always much smaller than the zeroth-order energy spacings).
	
	Suppose we start with one of the instantaneous eigenstates
	\begin{equation}
	\ket{\psi_{\mu}(\tau=0)}\equiv\ket{n_{\mu}(\lm(\tau=0))}
	\end{equation}
	with $\ket{n_{\mu}}$ an eigenstate belonging to some subspace spanned by eigenstates that share the same zeroth-order energy. Note that, in the following discussions in this subsection, we will use Greek indices ($\mu$, $\nu$, etc.) to denote eigenstates from this particular subspace and will use Latin indices (a,\,b,\,c, etc.) to denote those from other subspaces. Then, generally speaking, the state of the system at time $\tau$ can be expanded in terms of instantaneous eigenstates at that time
	\begin{equation}
	\begin{aligned}
	&\ket{\psi_{\mu}(\tau)}\\
	&=\sum_{\nu}U_{\mu\nu}(\tau)\ket{n_{\nu}(\lm(\tau))}
	+\sum_{a}U_{\mu a}(\tau)\ket{n_{a}(\lm(\tau))}
	\end{aligned}
	\end{equation}
	
	However, a generalization of (non-degenerate) adiabatic theorem implies that coefficients of $\ket{n_a}$'s (namely $U_{\mu a}$'s) can be significantly suppressed. This can be achieved once the inverse of characteristic time scale of the Hamiltonian (e.g., the frequency of a periodic Hamiltonian) is smaller when compared with the zeroth-order energy spacing. An outline of the proof is presented in Appendix A.
	
	Therefore, the expansion just becomes
	\begin{equation}
	\ket{\psi_{\mu}(\tau)}=\sum_{\nu}U_{\mu\nu}(\tau)\ket{n_{\nu}(\lm(\tau))}
	\label{SBJJ_Adiabtic_Ansatz}
	\end{equation}
	where one can readily show that $U_{\mu\nu}$ is a unitary matrix. Clearly, this means that we can neglect the contribution from states with different zeroth-order energies in our case. Substituting Eq.\,\eqref{SBJJ_Adiabtic_Ansatz} into the Schr\"{o}dinger equation (with $\hb=1$), we have
	\begin{equation}
	\begin{aligned}
	i\ket{\dot{\psi}_{\mu}}&=i\sum_{\nu}(\dot{U}_{\mu\nu}\ket{n_{\nu}}+U_{\mu\nu}\ket{\dot{n}_{\nu}})\\
	&=H\ket{\psi_{\mu}}=\sum_{\nu}U_{\mu\nu}E_{\nu}\ket{n_{\nu}}
	\end{aligned}
	\end{equation}
	Rearranging and taking overlap with $\bra{n_{\nu}}$, we have
	\begin{equation}
        \dot{U}_{\mu\nu}=-iU_{\mu\nu}E_{\nu}+i\sum_{\rho}U_{\mu\rho}(\mathcal{A}_i)_{\rho\nu}\dot{\lm}^i
	\end{equation}
	where
	\begin{equation}
	(\mathcal{A}_i)_{\mu\nu}\equiv i\bra{n_{\nu}}\pt_i\ket{n_{\mu}}
	\end{equation}
	is the non-Abelian Berry connection.
	
	Since all the $\ket{n_{\mu}}$ share the same zeroth-order energy, one is free to implement a shift of zero-point, canceling its contribution. Thus, the differential equation for $U_{\mu\nu}$ becomes
	\begin{equation}
        \dot{U}_{\mu\nu}=-iU_{\mu\nu}E^{(1)}_{\nu}+i\sum_{\rho}U_{\mu\rho}(\mathcal{A}_i)_{\rho\nu}\dot{\lm}^i
	\label{SBJJ_Ude}
	\end{equation} 
	
	\subsection{Instantaneous Eigenstates and the Time Evolution}
	To calculate the time-dependent evolution matrix $U_{\mu\nu}$, we need to derive the instantaneous eigenstates via perturbation theory. First, we define
	\begin{equation}
	\begin{aligned}
	&\chi_{n+}=\begin{pmatrix}
	\frac{1}{2}(1+\cos\theta)e^{-i\phi}\\
	\frac{1}{\sqrt{2}}\sin\theta\\
	\frac{1}{2}(1-\cos\theta)e^{i\phi}
	\end{pmatrix}\;,\;
	\chi_{n0}=\begin{pmatrix}
	-\frac{1}{\sqrt{2}}\sin\theta e^{-i\phi}\\
	\cos\theta\\
	\frac{1}{\sqrt{2}}\sin\theta e^{i\phi}
	\end{pmatrix}\;,\\
	&\chi_{n-}=\begin{pmatrix}
	\frac{1}{2}(1-\cos\theta)e^{-i\phi}\\
	-\frac{1}{\sqrt{2}}\sin\theta\\
	\frac{1}{2}(1+\cos\theta)e^{i\phi}
	\end{pmatrix}
	\end{aligned}
	\label{SBJJ_Eigenspinors_Def}
	\end{equation}
	which are eigenspinors of $\bm{s}\cdot\bm{n}$ corresponding to eigenvalue $+1$, $0$, and $-1$, respectively. We may also write spinors like $\chi_{zs} (s=0,\pm 1)$, which represent spinors in Eq.\,\eqref{SBJJ_Eigenspinors_Def} with $\bm{n}\rightarrow\bm{e}_z$.
	
	The unperturbed and perturbation Hamiltonians are
	\begin{equation}
	\mathcal{H}_0\equiv\begin{pmatrix}
	-\bm{s}\cdot\bm{n}(\tau) & \\
	& -s_z
	\end{pmatrix},\quad
	V\equiv\begin{pmatrix}
	& -k\\
	-k & 
	\end{pmatrix}
	\end{equation}
	We require that $k\ll 1$ [$k$ is defined in Eq. (9)] so that $V$ is a perturbation. Besides, we set $\omega\ll 1$ so that the (generalized) adiabatic theorem is valid. 
	
	Following the standard treatment of time-independent perturbation theory, one acquires the zeroth-order "good states" and the corresponding first-order energy correction
	\begin{equation}
	\begin{aligned}
	\Lambda^{(0)}_{+1}&=\begin{pmatrix}
	\frac{-1}{\sqrt{2}}\chi_{n+}\\
	\\
	\frac{e^{-i\phi}}{\sqrt{2}}\chi_{z+}
	\end{pmatrix},\quad E^{(1)}_+=\frac{1}{2}k(1+\cos\theta)\equiv|z|\\
	\\
	\Lambda^{(0)}_{+2}&=\begin{pmatrix}
	\frac{1}{\sqrt{2}}\chi_{n+}\\
	\\
	\frac{e^{-i\phi}}{\sqrt{2}}\chi_{z+}
	\end{pmatrix},\quad E^{(1)}_-=-\frac{1}{2}k(1+\cos\theta)=-|z|
	\end{aligned}
	\end{equation}
	Generally speaking, one may still need to calculate the first-order correction of eigenstates. However, the first-order ket is of order $O(k)$, as long as we restrict ourselves to cases for which
	\begin{equation}
	k,\;\omega\ll 1
	\label{SBJJ_cond}
	\end{equation}
	Contributions from these terms can be neglected, since the leading-order contribution from perturbation is of order $O(k/\omega)$. To show this, we employ Eq.\,\eqref{SBJJ_Ude}. For our case, the only component of Berry connection is $\mathcal{A}_i=\mathcal{A}_{\phi}$ with $\dot{\lm}^i=\dot{\phi}=\omega$. Therefore, the equation for $U$ is
    \begin{equation}
	\begin{aligned}
	\dot{U}_{11}&=-iU_{11}|z|+i(U\mathcal{A}_{\phi}\omega)_{11}\\
	\dot{U}_{12}&=iU_{12}|z|+i(U\mathcal{A}_{\phi}\omega)_{12}\\
	\dot{U}_{21}&=-iU_{21}|z|+i(U\mathcal{A}_{\phi}\omega)_{21}\\
	\dot{U}_{22}&=iU_{22}|z|+i(U\mathcal{A}_{\phi}\omega)_{22}\\
	\end{aligned}
	\end{equation}
   Or equivalently,
	\begin{equation}
	\frac{\rmd}{\rmd\phi}U=iU(\mathcal{A}_{\phi}-\frac{|z|}{\omega}\sg_3)
	\label{SBJJ_Udynamics}
	\end{equation}
	The dynamical effect is of order $O(|z|/\omega)\sim O(k/\omega)$, as promised.
	
	Therefore, one can easily deduce that the leading-order of non-Abelian connection is
	\begin{equation}
	\begin{aligned}
	&\begin{cases}
	\;(\mathcal{A}_{\phi})_{11}=(\mathcal{A}_{\phi})_{22}=\frac{1}{2}(1+\cos\theta)&\\
	\;(\mathcal{A}_{\phi})_{12}=(\mathcal{A}_{\phi})_{21}=\frac{1}{2}(1-\cos\theta)&
	\end{cases}\\
	&\Leftrightarrow
	\mathcal{A}_{\phi}=\frac{1}{2}(1+\cos\theta)+\frac{1}{2}(1-\cos\theta)\sg_1
	\end{aligned}
	\end{equation}
	So Eq.\,\eqref{SBJJ_Udynamics} now becomes
	\begin{equation}
	\begin{aligned}
	&\frac{\rmd U^{\dg}}{\rmd\phi}\\
	&=-\frac{i}{2}(1+\cos\theta)U^{\dg}\\
	&-i\left(\frac{1}{2}(1-\cos\theta)\sg_1-\frac{k}{2\omega}(1+\cos\theta)\sg_3\right)U^{\dg}
	\end{aligned}
	\label{SBJJ_EOMU_1}
	\end{equation}
	with the initial condition: $U^{\dg}(\phi=0)=U(\phi=0)=1$. 
	
	The first term of the equation above only provides an unimportant phase factor and can be neglected. Integrating the equation, we have
	\begin{equation}
	U(\phi)=\begin{pmatrix}
	\cos\frac{\alpha}{2}+i\sin\frac{\alpha}{2}\cos\Theta & i\sin\frac{\alpha}{2}\sin\Theta\\
	& \\
	i\sin\frac{\alpha}{2}\sin\Theta & \cos\frac{\alpha}{2}-i\sin\frac{\alpha}{2}\cos\Theta
	\end{pmatrix}
	\label{SBJJ_Uresult}
	\end{equation}
	where
	\begin{equation}
	\begin{aligned}
	&\frac{\alpha}{2}\equiv\frac{1}{2}\phi(1-\cos\theta)\sqrt{1+\left(\eta\frac{k}{\omega}\right)^2},\\ 
	&\Theta\equiv\pi-\arctan\left(\frac{\omega}{\eta k}\right),\quad\eta\equiv\frac{1+\cos\theta}{1-\cos\theta}
	\end{aligned}
	\end{equation}

\subsection{Particle Current}
	With the evolution matrix, we can derive the time evolution of the particle number difference between two condensates
	\begin{equation}
	\Delta N(\tau)\equiv\frac{N_2(\tau)-N_1(\tau)}{N_1(\tau)+N_2(\tau)}=N_2(\tau)-N_1(\tau)
	\end{equation}
	Note that we have normalized the total particle number to unity. 
 
 We start with, for example, the initial state
	\begin{equation}
	\ket{\psi(\tau=0)}=
	\begin{pmatrix}
	\frac{1}{\sqrt{2}}\chi_{n+}\\
	\\
	\frac{1}{\sqrt{2}}\chi_{z+}
	\end{pmatrix}=\ket{\Lambda^{(0)}_{+2}(\phi=0)}
	\label{ini_cond}
	\end{equation}
	Then, the state after evolution is
	\begin{equation}
	\begin{aligned}
	&\ket{\psi(\tau)}\\
	&=U_{21}(\phi)\ket{\Lambda^{(0)}_{+1}(\phi)}+U_{22}(\phi)\ket{\Lambda^{(0)}_{+2}(\phi)}\\
	&=\begin{pmatrix}
	\frac{1}{\sqrt{2}}\chi_{n+}(-i\sin\frac{\alpha}{2}\sin\Theta+\cos\frac{\alpha}{2}-i\sin\frac{\alpha}{2}\cos\Theta)\\
	\\
	\frac{e^{-i\phi}}{\sqrt{2}}\chi_{z+}(i\sin\frac{\alpha}{2}\sin\Theta+\cos\frac{\alpha}{2}-i\sin\frac{\alpha}{2}\cos\Theta)
	\end{pmatrix}
	\end{aligned}
	\end{equation}
	where $\phi=\omega\tau$.

	Therefore, one can calculate the particle number difference after an arbitrary evolution
	\begin{equation}
	\begin{aligned}
	\Delta N
 &=-\sin2\Theta\sin^2\frac{\alpha}{2}\\
	&=\sin\left(2\arctan\left(\frac{\omega}{\eta k}\right)\right)\\
	&\times\sin^2\left(\frac{1}{2}\omega(1-\cos\theta)\sqrt{1+\left(\eta\frac{k}{\omega}\right)^2}\tau\right)
	\label{SBJJ_PREDICTION}
	\end{aligned}
	\end{equation}
 which, in terms of the original dimensionful variables ($\Omega$, $t$ and $K$), is
 \begin{equation}
\begin{aligned}
    \Delta N(t)&=\sin\left(2\arctan\left(\frac{\hb\Omega}{\eta K}\right)\right)\\
	&\times\sin^2\left(\frac{1}{2}(1-\cos\theta)\sqrt{1+\left(\frac{\eta K}{\hb\Omega}\right)^2}\Omega t\right)
\end{aligned}
\label{dN_LF_SI}
 \end{equation}
	
	\subsection{Discussion}
	From Eq.\,\eqref{SBJJ_PREDICTION}, the maximum particle difference is given by
	\begin{equation}
	\Delta N_{\mathrm{max}}=\sin\left\{2\arctan\left(\frac{\omega}{k}\frac{1-\cos\theta}{1+\cos\theta}\right)\right\}
	\end{equation}
	The maximum value of the amplitude of $\Delta N$ can be reached if
	\begin{equation}
		2\arctan\left(\frac{\omega}{k}\frac{1-\cos\theta}{1+\cos\theta}\right)=\frac{\pi}{2}
		\label{SBJJ_maximum_cond}
	\end{equation}
	
	Besides, one can consider the limit case for which $\eta k/\omega\ll 1$. In this case, we have
	\begin{equation}
        \begin{aligned}
        \Delta N&\simeq 2\eta\frac{k}{\omega}\sin^2\left(\frac{1}{2}\omega(1-\cos\theta)\tau\right)\\
        &=2\left(\frac{1+\cos\theta}{1-\cos\theta}\right)\frac{K}{\hb\Omega}\sin^2\left(\frac{1}{2}(1-\cos\theta)\Omega t\right)
        \end{aligned}
        \label{dN_approx}
	\end{equation}
	
	The particle current is given by the time derivative of $\Delta N$:
	\begin{equation}
	\begin{aligned}
	J&\equiv\pt_{t}\Delta N\\
	&=\frac{K}{\hb}(1+\cos\theta)\sin((1-\cos\theta)\Omega t)\\
	&\propto\sin\left((1-\cos\theta)\Omega t\right)
	\end{aligned}
	\label{Intuition}
	\end{equation}

	\begin{figure}[htbp]
		\centering
		\includegraphics[width=0.48\textwidth]{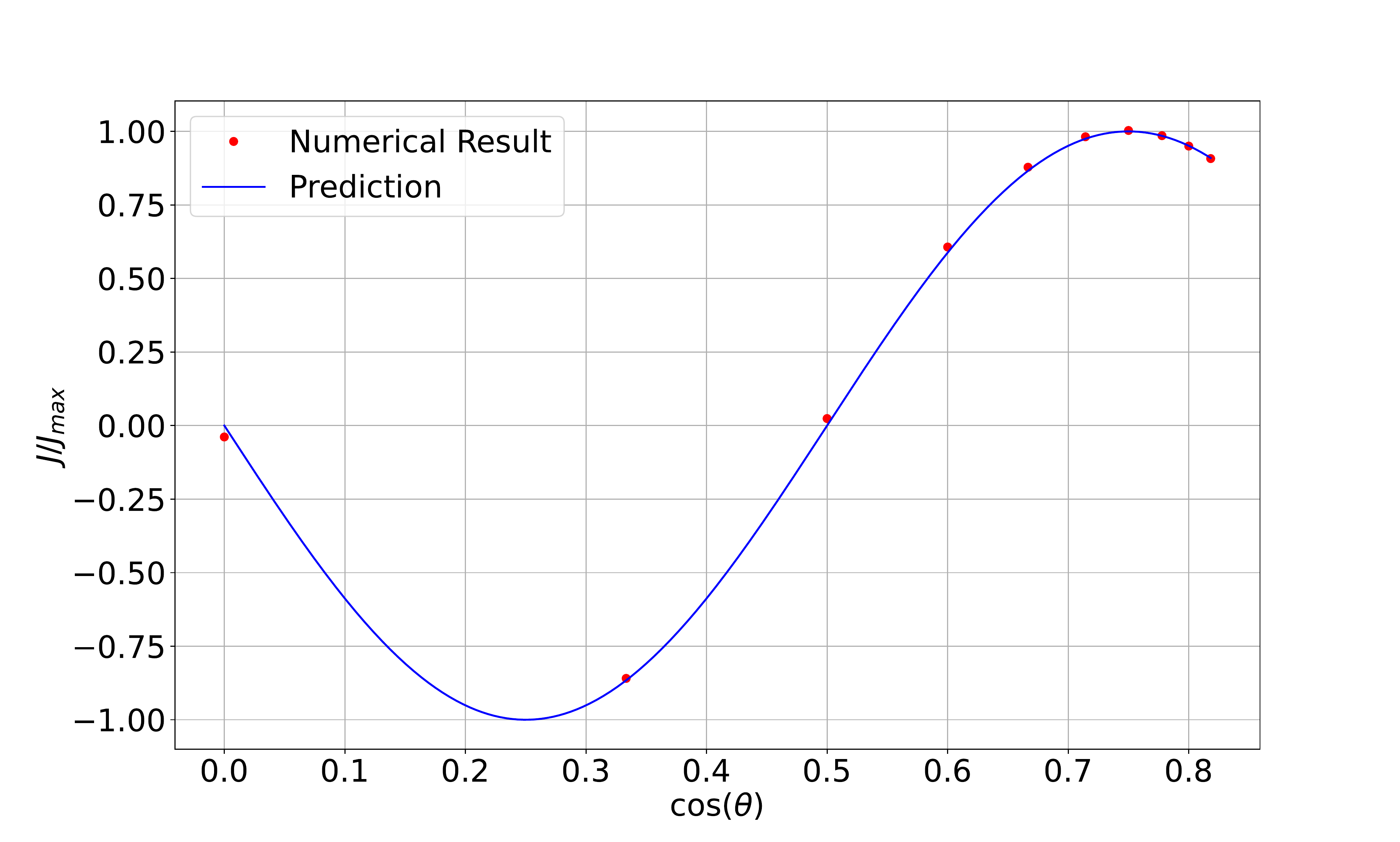}
		\caption{$J-\theta$ relation of Eq.\,\eqref{J-N} given $N=1$, where $J_{max}$ is given by Eq.\,\eqref{Intuition}.}
		\label{J-theta}
	\end{figure}
	On the other hand, consider a spin-1 particle in the presence of a rotating magnetic field $\bm{B}_1(t)=B\bm{n}(t)$ (having already appeared in Sec.\,\ref{GPE}). Just like our BJJ system, we assume that the spin stays in its $s_z=+1$ eigenstate
	\begin{equation}
	\chi_{n+}=\begin{pmatrix}
	\frac{1}{2}(1+\cos\theta)e^{-i\phi}\\
	\frac{1}{\sqrt{2}}\sin\theta\\
	\frac{1}{2}(1-\cos\theta)e^{i\phi}
	\end{pmatrix}
	\end{equation} 
	So we can calculate its corresponding (Abelian) Berry connection\cite{NQ_berryphase}
	\begin{equation}
	\mathcal{A}_{\phi}=i\bra{\chi_{n+}}\pt_{\phi}\ket{\chi_{n+}}=\cos\theta
	\end{equation}
	and the Berry phase accumulated after one cycle's adiabatic evolution
	\begin{equation}
	\gamma=\oint_{\mathcal{C}}\rmd\bm{r}\cdot\bm{\mathcal{A}}(\bm{r})=\int_0^{2\pi}\rmd\phi\mathcal{A}_{\phi}=2\pi\cos\theta
	\end{equation}
	
	Now, we have an intuitive interpretation for Eq.\,\eqref{Intuition}. After exactly $N$ cycles' evolution, the "current" (i.e. the derivative of particle number) $J$ is proportional to
	\begin{equation}
	J\propto\sin\left(2N\pi(1-\cos\theta)\right)=\sin\left(-N\cdot 2\pi\cos\theta\right)
        \label{J-N}
	\end{equation}
	So the current is proportional to the sine of phase difference of the BJJ system, as is shown in Fig.\,\ref{J-theta}. The change of phase difference comes from the collective adiabatic rotation of spins in the left side of the BJJ\,(the minus sign arises because the phase difference is defined as the phase of the right side minus the left side and spins living on the right do not change their phase). This coincides with the intuitive argument given in Sec.\,\ref{Intro}. Besides, from Eq.\,\eqref{dN_LF_SI}, \eqref{dN_approx} and \eqref{Intuition} one may notice that the oscillation of $\Delta N$ and $J$ have no dependence on the value of external magnetic field. This can also be explained by the fact that the value of geometrical phase does not depend on the strength of external field, but on the area enclosed by the closed loop in parameter space.

\begin{figure}[htbp]
		\subfloat[Obtaining rotating field by treating two components separately.]
		{  \label{SCHEME_2}
		   \includegraphics[width=.4\textwidth]
		   {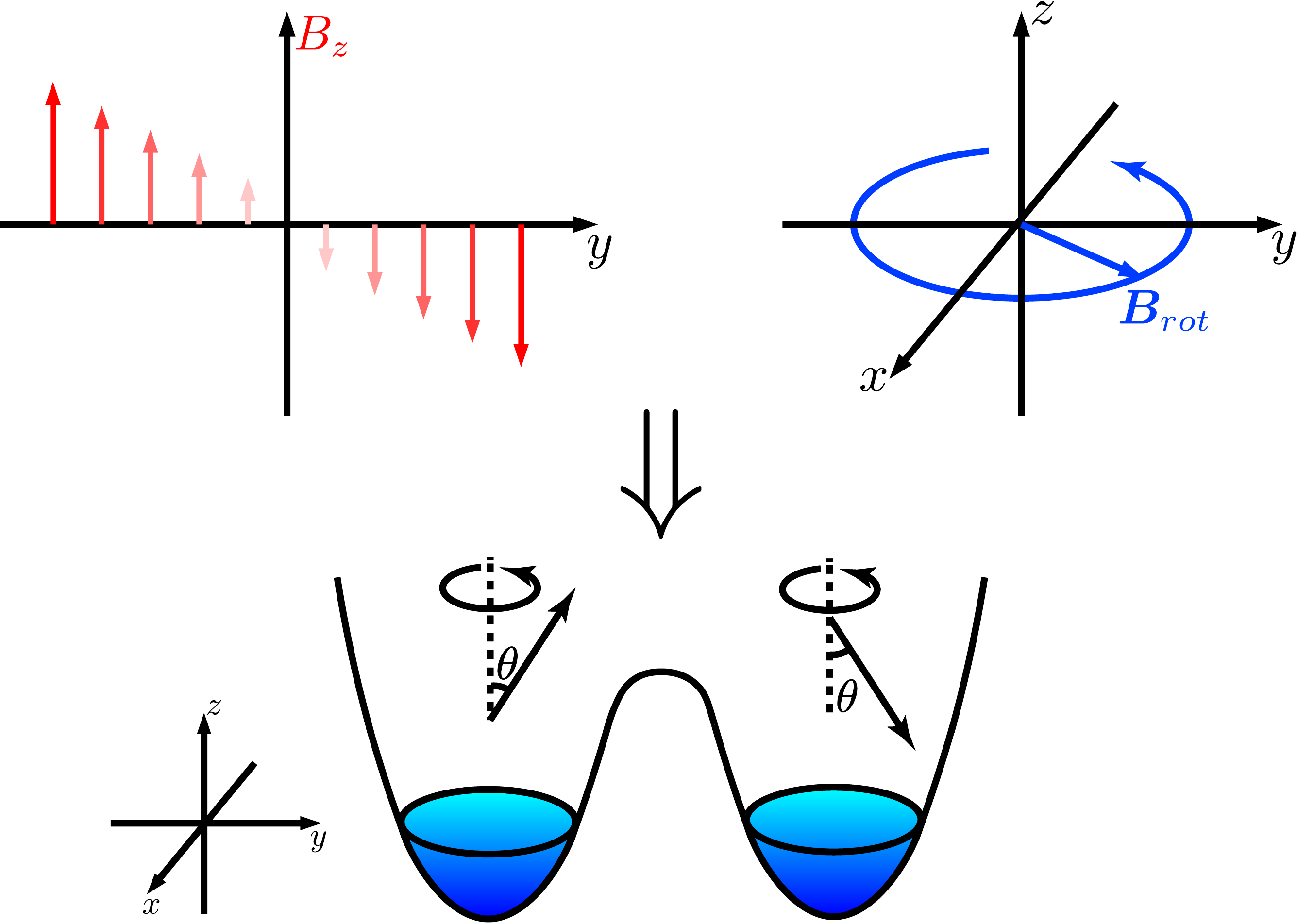}
		}

		\subfloat[Particle current obtained by numerical calculation. Parameter setting: $(k,\omega,\theta)=(5\times10^{-4},10^{-2},\arccos(1/3))$.]
		{  \label{EXP}
		   \includegraphics[width=.4\textwidth]
		   {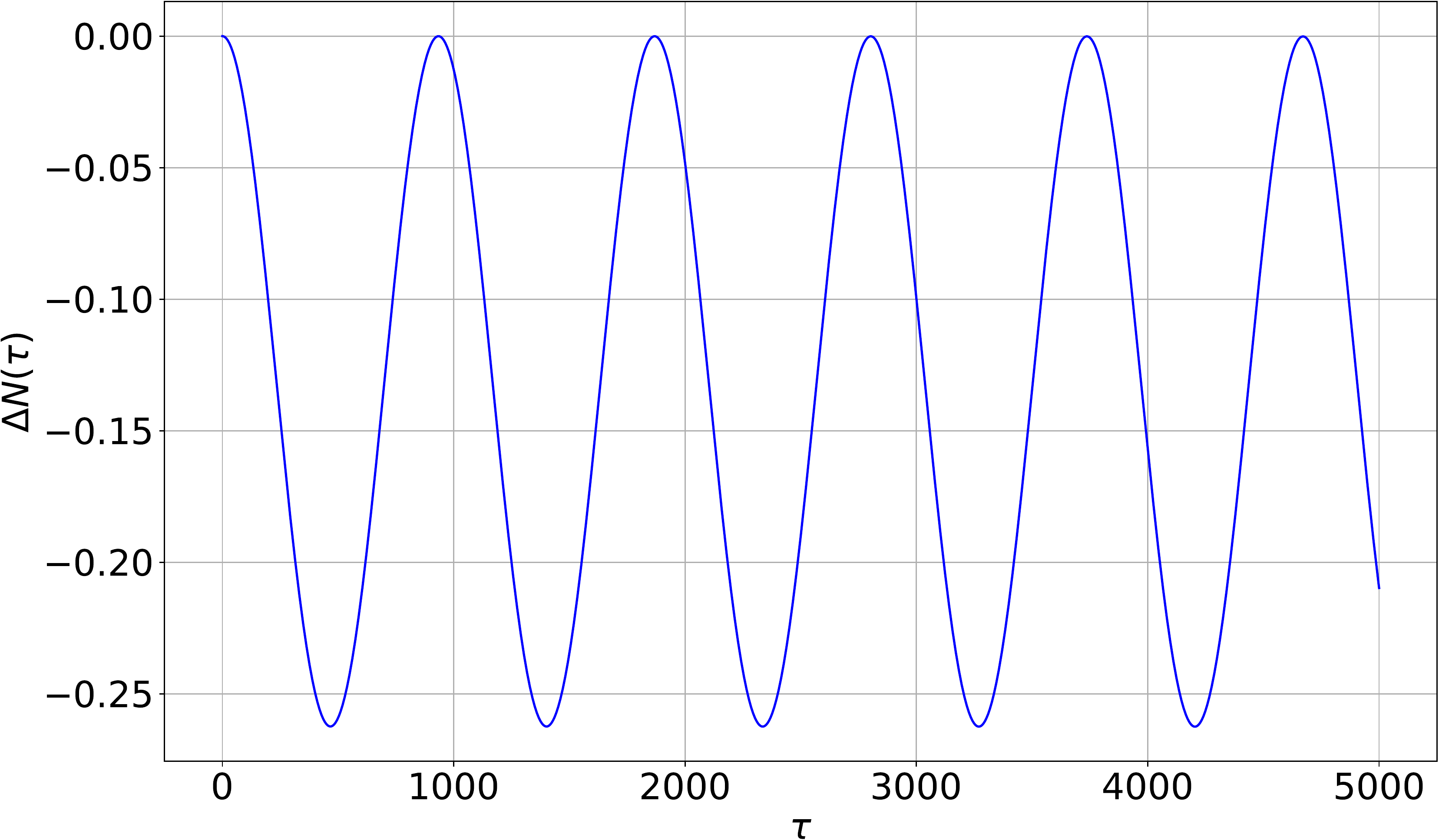}
		}
	 
		\caption{Compared to the schematic demonstration in Fig. 1, here we propose an experimentally feasible method to independently tune magnetic fields in two traps.}
		\label{SCHEMES}
	 \end{figure}

 	Based on this intuitive argument, we propose a scheme to independently vary the magnetic fields on two traps. By introducing a geometrical phase difference, we can induce the particle current without the need for a static field on the right side. Instead, we rotate the field on the right side at a different polar angle. This can be achieved in two steps, as illustrated in Fig.\, \ref{SCHEME_2}. First, we generate a field gradient along the system to provide static, opposite $B_z$ components to atoms on different sides. Then, we uniformly apply a circularly polarized beam, resulting in both sides having the same rotating $B_x/B_y$ components. This scheme has been verified through straightforward numerical calculations (see Fig.\,\ref{EXP}), which also demonstrate its ability to produce current. By treating the static and rotating components separately, this scheme may be more applicable in experimental settings.

	\section{The High-Frequency Limit}
	\label{HF}
	
	\subsection{The Floquet Hamiltonian and Particle Current}
	In this section we shall consider the high-frequency limit, where the frequency $\omega\gg 1$ is much larger than the Zeeman energy. In that case, the time-averaged property of the system is captured by a stroboscopic time-independent Floquet Hamiltonian $\mathcal{H}_F[\tau_0]$, defined via the evolution operator in one period\cite{Floquet_ref}
	\begin{equation}
	\begin{aligned}
	&U(\tau_0+T,\tau_0)
	=\mathcal{T}\exp\left(-i\int_{\tau_0}^{\tau_0+T}\rmd t \mathcal{H}(t)\right)\\
	&\equiv\exp\left(-i\mathcal{H}_F[\tau_0]T\right)
	\end{aligned}
	\end{equation}
	where $\mathcal{H}(t)$ is the original time-dependent Hamiltonian and $\mathcal{T}$ denotes the time ordering. By definition, the form of $\mathcal{H}_F[\tau_0]$ depends on our choice of $\tau_0$, which is the beginning of the stroboscopic period. This is called the Floquet gauge choice. In what follows, we may simply fix this gauge by setting $\tau_0=0$ and hence omit its symbol.
	
	Once the frequency $\omega\gg 1$ is sufficiently large, we can find the approximate form of $\mathcal{H}_F$ systematically by expanding it as a series of $1/\omega$, which is known as the Magnus expansion \cite{Floquet_ref}:
	\begin{equation}
		\mathcal{H}_F=\sum_{n=0}^{\infty}\mathcal{H}_F^{(n)},\; \mathcal{H}_F^{(n)}\sim O(\omega^{-n})
	\end{equation}  
	A few leading-order terms of the expansion (with the gauge choice $\tau_0=0$) are given by
	\begin{equation}\label{magnus_expansion}
	\begin{aligned}
	&\mathcal{H}_F^{(0)}=\mathcal{H}_0\\
	&\mathcal{H}_F^{(1)}=\frac{1}{\omega}\sum_{l=1}^{\infty}\frac{1}{l}
	\left([\mathcal{H}_l,\mathcal{H}_{-l}]-[\mathcal{H}_l,\mathcal{H}_0]+[\mathcal{H}_{-l},\mathcal{H}_0]\right)\\
	&\cdots
	\end{aligned}
	\end{equation}
	where $\mathcal{H}_l$ is the Fourier transformation of $\mathcal{H}(\tau)$:
	\begin{equation}\label{magnus_ft}
	\mathcal{H}(\tau)=\sum_l\mathcal{H}_le^{il\omega\tau}\;\leftrightarrow\;
	\mathcal{H}_l=\frac{1}{T}\int_{0}^{T}\rmd\tau e^{-il\omega\tau}\mathcal{H}(\tau)
	\end{equation}
	
	In this system, we have 
	\begin{equation}\label{magnus_H}
	\begin{aligned}
	&\mathcal{H}(\tau)=
	\begin{pmatrix}
	-\bm{s}\cdot\bm{n}(\tau) & -k\\
	-k & -s_z
	\end{pmatrix},\;\\
	&\bm{n}(\tau)=(\sin\theta\cos\omega\tau,\sin\theta\sin\omega\tau,\cos\theta)^T
	\end{aligned} 
	\end{equation}
	Thus
	\begin{equation}
	\mathcal{H}_F^{(0)}=\frac{1}{T}\int_{0}^{T}\rmd\tau\mathcal{H}(\tau)
	=\begin{pmatrix}
	-s_z\cos\theta & -k\\
	-k & -s_z
	\end{pmatrix}
	\end{equation}
	The upper left element is changed from $-\bm{s}\cdot\bm{n}$ to $-s_z\cos\theta$. Heuristically, this new term denotes the average Zeeman energy felt by particles when the frequency is infinitely large.
	
\begin{figure}[tbp]
		\centering
		\includegraphics[width=0.45\textwidth]{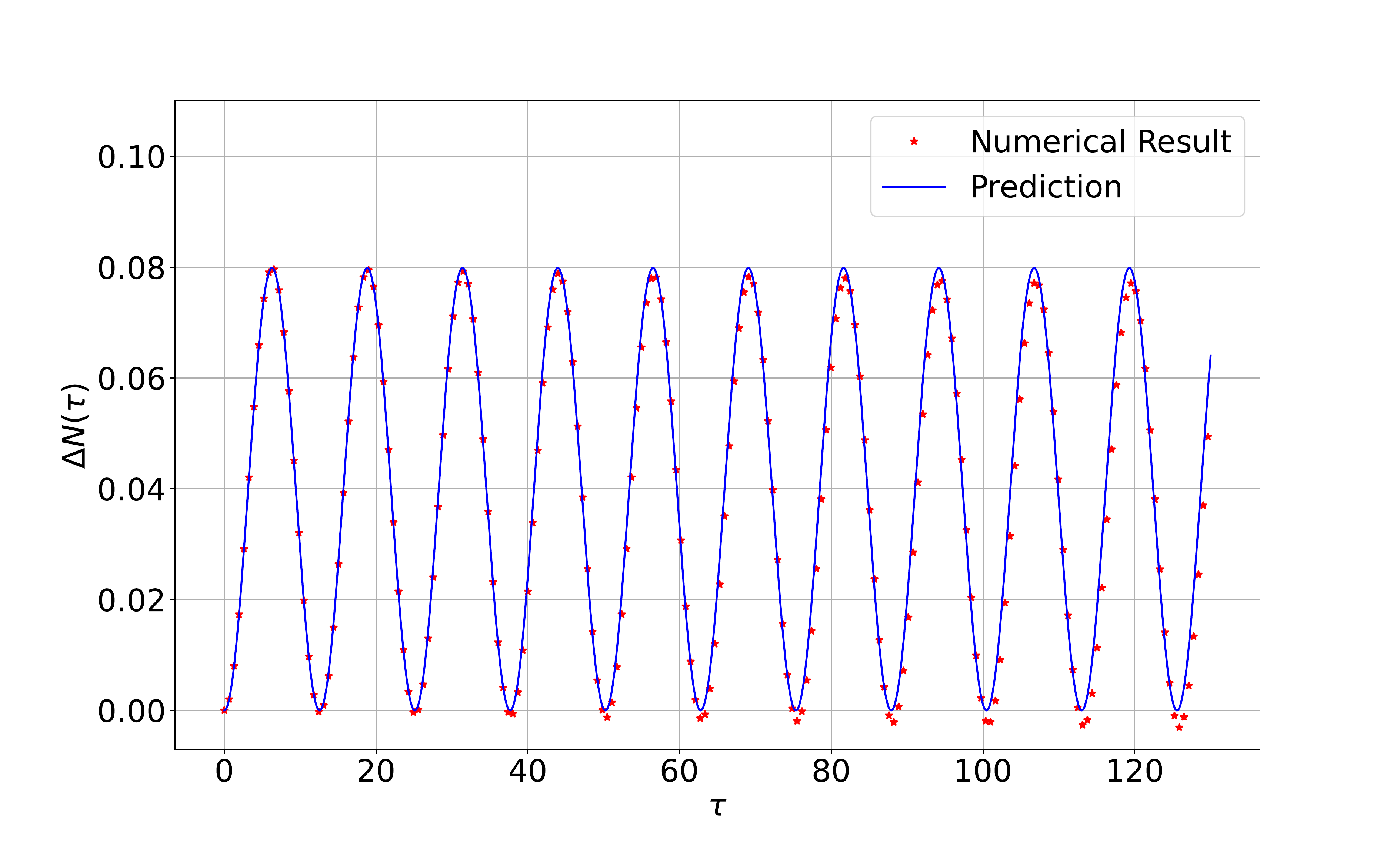}
		\caption{Numerical result in high-frequency limit with theoretical prediction using zeroth-order Floquet Hamiltonian. Parameter setting: $(k,\omega,\theta)=(10^{-2},300,\pi/3),\;\Lambda_+=(1/\sqrt{2})(1,1)^T,\;\Lambda_0=\Lambda_-=0$}
		\label{high_frequency_expansion}
	\end{figure}

	In fact, $\mathcal{H}_F^{(0)}$ is sufficient to describe the high-frequency property of the system, so there is no need to calculate higher order terms. Fig.\,\ref{high_frequency_expansion} shows the predicted particle number difference together with numerical result. The agreement of these two curves implies that the zeroth-order term in Magnus expansion is enough for quantitative discussions. 
	
	The argument above only relies on the absolute value of $\omega$. Thus $\mathcal{H}_F$ (and hence the dynamical evolution) will not change if we reverse the rotation of magnetic field ($\omega \rightarrow -\omega$). This is completely different from the low-frequency case in which $\Delta N$ flips sign when the rotation is reversed, as Eq.\,\eqref{SBJJ_PREDICTION} implies. 
	
	Clearly, a change of basis
	\begin{equation}
	\begin{aligned}
	&\Lambda=\begin{pmatrix}
	\Lambda_1\\
	\Lambda_2
	\end{pmatrix}
	\:,\:
	\Lambda_i=\begin{pmatrix}
	\Lambda_{i,+}\\
	\Lambda_{i,0}\\
	\Lambda_{i,-}
	\end{pmatrix}\:(i=1,2)\\
	\;
	\Rightarrow
	\;
	&\tilde{\Lambda}=\begin{pmatrix}
	\Lambda_+\\
	\Lambda_0\\
	\Lambda_-
	\end{pmatrix}
	\:,\:
	\Lambda_\sg=\begin{pmatrix}
	\Lambda_{1,\sg}\\
	\Lambda_{2,\sg}
	\end{pmatrix}\:(\sg=+,0,-)
	\end{aligned}
	\end{equation}
	renders the Floquet Hamiltonian block-diagonal:
	\begin{equation}
	\begin{aligned}
	&\mathcal{H}_F=\mathrm{diag}\left(\mathcal{H}_{F,+},\:\mathcal{H}_{F,0},\:\mathcal{H}_{F,-}\right)\\
	&\mathcal{H}_{F,+}=-
	\begin{pmatrix}
	\cos\theta & k\\
	k & 1
	\end{pmatrix},\quad
	\mathcal{H}_{F,0}=-
	\begin{pmatrix}
	0 & k\\
	k & 0
	\end{pmatrix},\\
	&\mathcal{H}_{F,-}=-
	\begin{pmatrix}
	-\cos\theta & k\\
	k & -1
	\end{pmatrix}
	\end{aligned}
	\end{equation} 
	Thus we conclude that, in the high-frequency limit, the dynamics of three spin degrees of freedom (labeled by $\sg=+1,0,-1$) are decoupled. As a result, one may study these three Hamiltonians separately.
	
	\subsection{Discussion}
	The dynamical evolution of three internal states are governed by three $2\times 2$ matrices:
	\begin{eqnarray}
	\mathcal{H}_{F,+}&=&-\mathcal{H}_{F,-}\nonumber\\
	&=&-\frac{1}{2}(1+\cos\theta)-k\sg_x+\frac{1}{2}(1-\cos\theta)\sg_z\nonumber\\\\
	\mathcal{H}_{F,0}&=&-k\sg_x 
	\end{eqnarray} 
	and the corresponding frequencies of oscillation are then given by Bohr frequencies
	\begin{equation}
	\omega_+=\omega_-=2\sqrt{\frac{1}{4}(1-\cos\theta)^2+k^2},\quad\omega_0=2k
        \label{bohr_freqs}
	\end{equation} 
    namely (in terms of dimensionful frequencies given by $\omega\tau\rightarrow \Omega t$):
    \begin{equation}
      \begin{aligned}
        &\Omega_+=\Omega_-=\frac{1}{\hb}\sqrt{\left(g\mu_B B(1-\cos\theta)\right)^2+(2K)^2}\\
        &\Omega_0=\frac{2K}{\hb}
      \end{aligned}     
    \end{equation}
	 Notice that frequencies from $s_z=\pm 1$ are identical. In practice, one may distinguish the oscillation of these two levels by adding a small energy difference. For example:
	\begin{equation}
	\begin{aligned}
	&\mathcal{H}(\tau)=
	\begin{pmatrix}
	-\bm{s}\cdot\bm{n}(\tau) & -k\\
	-k & -s_z
	\end{pmatrix}\\
	\;\rightarrow\;
	&\mathcal{H}'(\tau)=
	\begin{pmatrix}
	-\bm{s}\cdot\bm{n}(\tau)+\delta & -k\\
	-k & -s_z-\delta
	\end{pmatrix}
	\end{aligned}
        \label{adding_energy_difference}
	\end{equation}
	and frequencies become
	\begin{equation}
	\begin{aligned}
	&\omega'_+=2\sqrt{\left[\frac{1}{2}(1-\cos\theta)+\delta\right]^2+k^2}\\
	&\omega'_0=2\sqrt{\delta^2+k^2}\\
	&\omega'_-=2\sqrt{\left[\frac{1}{2}(1-\cos\theta)-\delta\right]^2+k^2}\\
	\end{aligned}
        \label{energy_difference_effect}
	\end{equation}
	which are three distinct values.

	\begin{figure}[htbp]
		\centering
		\includegraphics[width=0.48\textwidth]{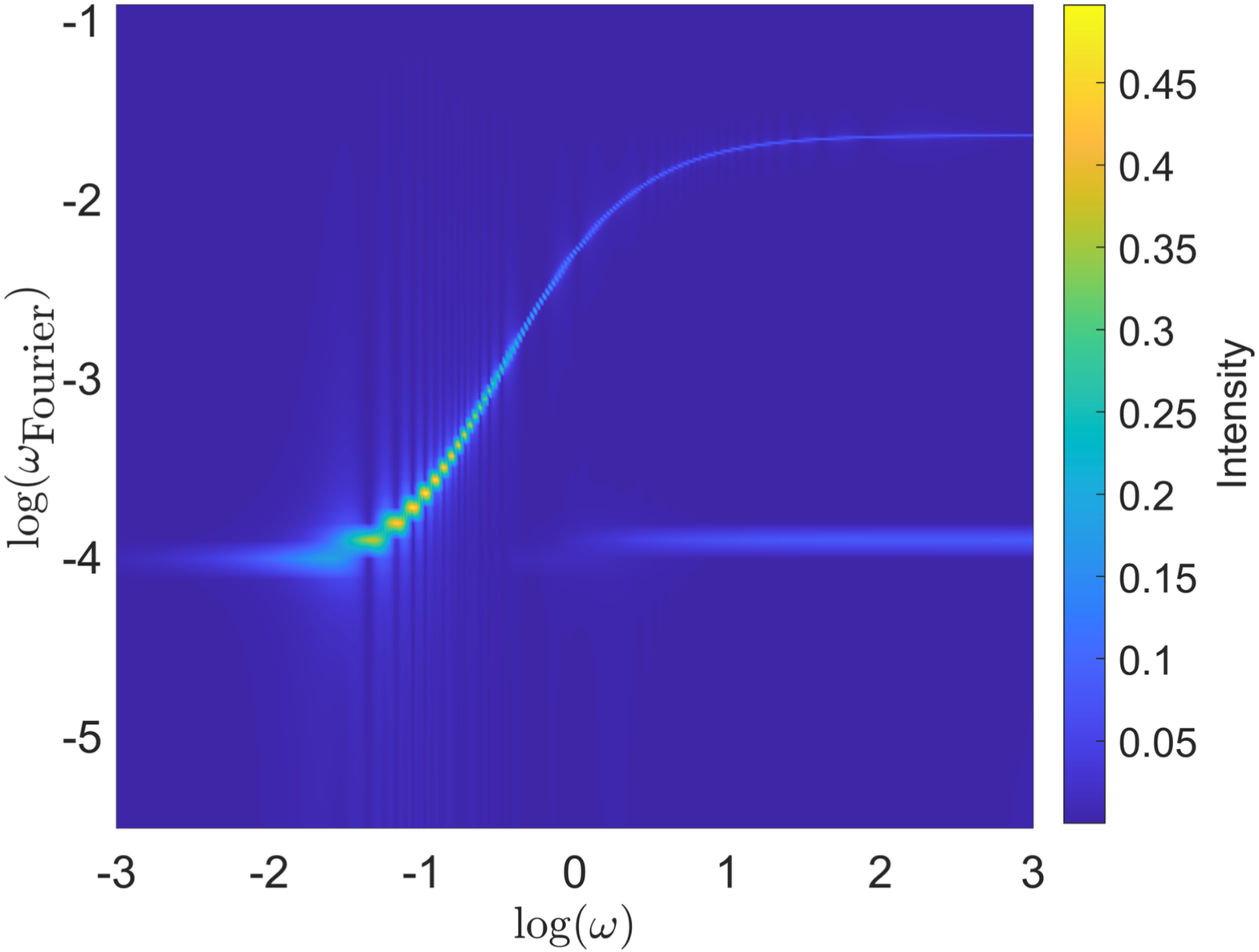}
		\caption{Fourier transformation of $
		\Delta N(\tau)$. The color indicates the intensity of the Fourier component $\Delta N(\omega_\text{Fourier})$, and $\omega$ represents the rotation frequency of the external magnetic field. Parameter setting: $(k,\theta)=(10^{-2},\pi/5)$.\;Initial condition [c.f. Eq.\,\eqref{ini_cond}]: $(1/\sqrt{2})(\chi_{n+},\chi_{z+})$.}
		\label{colorfig}
	\end{figure}
	Presently, we do not clearly understand the system's behavior when frequency $\omega$ is neither too low nor too high. In that case, it turns out that the oscillation behavior of $\Delta N$ is no longer periodic. This can be seen from the numeric Fourier transformation of $\Delta N(\tau)$, as is shown in Fig.\,\ref{colorfig}. One can observe sharp peak(s) of the Fourier spectrum, provided that $\omega \ll 1$ or $\omega \gg 1$. On the contrary, when $\omega\sim O(1)$ the spectrum is continuous, signaling the nonperiodicity.
	
	This nonperiodicity may be explained in terms of higher-order terms of the Magnus expansion. To see this, notice that the characteristic oscillation frequencies of different expansion terms are, in general, different. For instance, by employing Eqs.\,\eqref{magnus_expansion},\,\eqref{magnus_ft}, and \eqref{magnus_H}, we can calculate the first-order correction of the Floquet Hamiltonian
	\begin{equation}
	\mathcal{H}_F^{(1)}=\frac{1}{\omega}
	\begin{pmatrix}
	-\sin\theta\left(s_x\cos\theta+\frac{1}{2}s_z\sin\theta\right) & iks_y\sin\theta \\
	 & \\
	-iks_y\sin\theta & 0
	\end{pmatrix}
	\end{equation}
	The eigenvalue of $\mathcal{H}_F^{(1)}$ is
	\begin{equation}
	\left\{0,\,\pm\frac{k\sin\theta}{\omega},\,\pm\frac{\sin\theta}{2\sqrt{2}\omega}\sqrt{5+3\cos2\theta+8k^2}\right\}
	\end{equation}
	so its characteristic frequency is indeed different from that of $\mathcal{H}_F^{(0)}$. When $\omega$ is in high-frequency regime (\textit{i.e.} $\omega\gg 1$), the contribution from $\mathcal{H}_F^{(1)}$ is suppressed by a factor of order $1/\omega$. When $\omega$ becomes lower, this contribution becomes significant, yielding peaks with different frequencies. Clearly, this argument can be generalized to higher-order terms: more and more frequencies become significant and the oscillation becomes non-periodic at last.

 	\section{The Effect of Self-Interaction}
	\label{Interaction}
	In this section we discuss the effect of weak self-interaction to our results. To estimate the interaction strength, in this section we normalize the macroscopic wavefunction and $\Lambda_{1/2}$, such that
	\begin{equation}
		\begin{aligned}
			&\int\;\rmd^3r\Psi^{\dg}(\bm{r},t)\Psi(\bm{r},t)=N_{tot}\\
			&\Lambda^{\dg}_1\Lambda_1=N_1,\;\:\Lambda^{\dg}_2\Lambda_2=N_2
		\end{aligned}
	\end{equation}
	where $N_{tot}$ is the total particle number and $N_{1/2}$ is the particle number on the left/right side.
	 The effect of interaction can be encapsulated in nonlinear terms \cite{internalBJJ_2} of the time-dependent GPE:
	\begin{equation}
		\begin{aligned}
		&i\hb\pt_t\Psi\\
		&=\left(
		-\frac{\hb^2}{2m}\nabla^2
		+V_{\mathrm{ext}}(\bm{r})
		-g\mu_B\bm{s}\cdot\bm{B}(\bm{r},t)\right)\Psi\\
		&+\alpha|\Psi|^2\Psi
		+\beta\left(\Psi^{\dg}\bm{s}\Psi\right)\cdot\bm{s}\Psi
		\end{aligned}
		\label{INT_GPE}
		\end{equation}
	where $\alpha$ and $\beta$ denote the strength of spin-independent and spin-dependent interactions \cite{spin-dep_int_1,spin-dep_int_2}, respectively. To simplify the question, we use the same two-mode approximation (Eq.\,\eqref{ansatz}), which yields
	\begin{equation}
	\begin{cases}
	\;i\pt_{\tau}\Lambda_1=-\bm{s}\cdot\bm{n}(\tau)\Lambda_1-k\Lambda_2+uN_1\Lambda_1+v\bm{s}_1\cdot\bm{s}\Lambda_1&\\
	\;i\pt_{\tau}\Lambda_2=-k\Lambda_1-s_z\Lambda_2+uN_2\Lambda_2+v\bm{s}_2\cdot\bm{s}\Lambda_2&
	\end{cases}
	\end{equation}
	with the definition
	\begin{equation}
	\begin{aligned}
	&u\equiv\frac{1}{g\mu_BB}\alpha\int\rmd^3r\phi_i^4,\;\:v\equiv\frac{1}{g\mu_BB}\beta\int\rmd^3r\phi_i^4\\
	&\bm{s}_i\equiv\Lambda^{\dg}_i\bm{s}\Lambda_i\;\;(i=1,2)
	\end{aligned}
	\end{equation}
	Notice that we have assumed that the shape of $\phi_1$ and $\phi_2$ are identical, for simplicity. 

\begin{figure}[tb]
		\subfloat[Spin-independent interaction]
		{  \label{INT_u}
		   \includegraphics[width=.45\textwidth]
		   {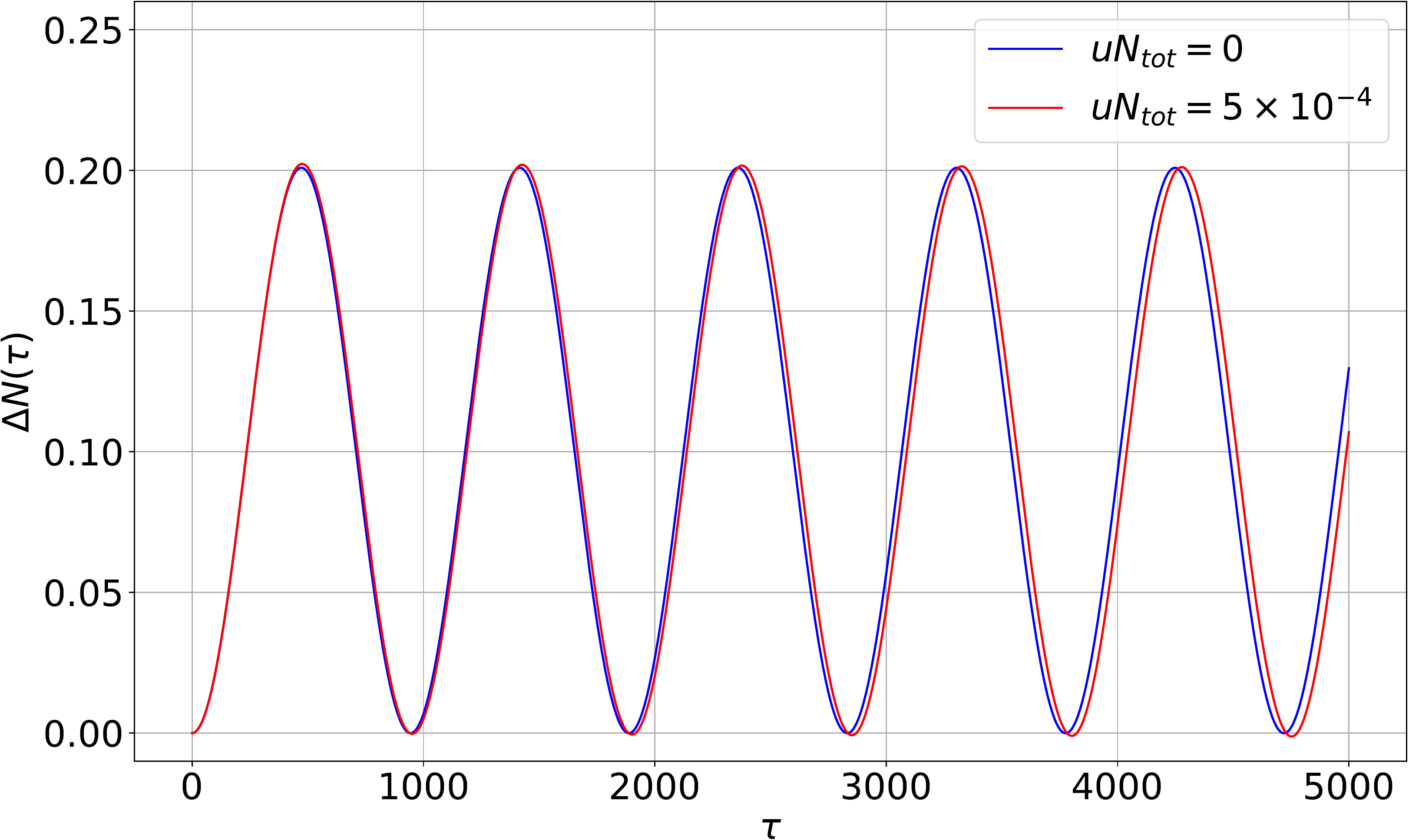}
		}

		\subfloat[Constant $\Delta E$]
		{  \label{INT_dE}
		   \includegraphics[width=.45\textwidth]
		   {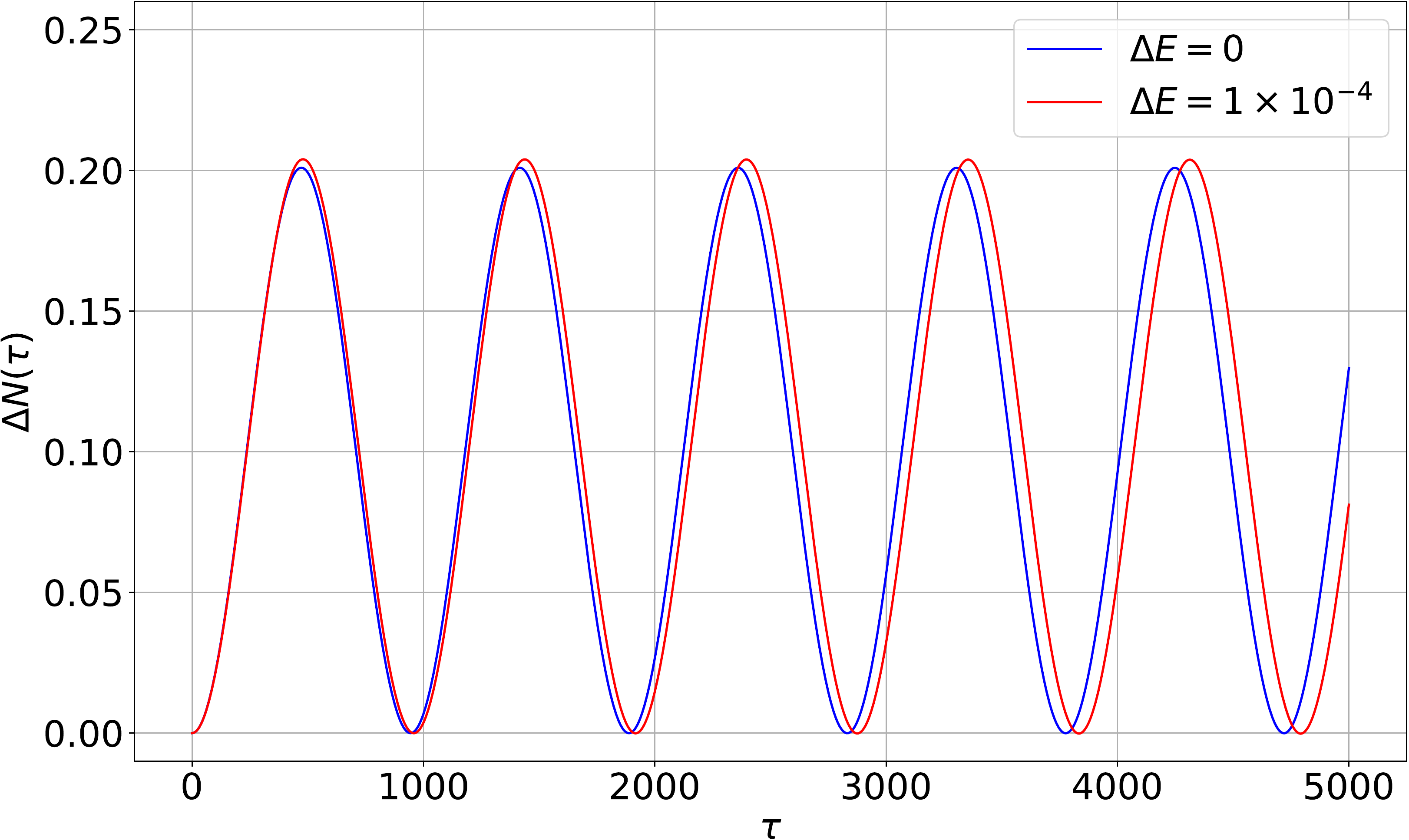}
		}

		\subfloat[Spin-dependent interaction]
		{  \label{INT_v}
		   \includegraphics[width=.45\textwidth]
		   {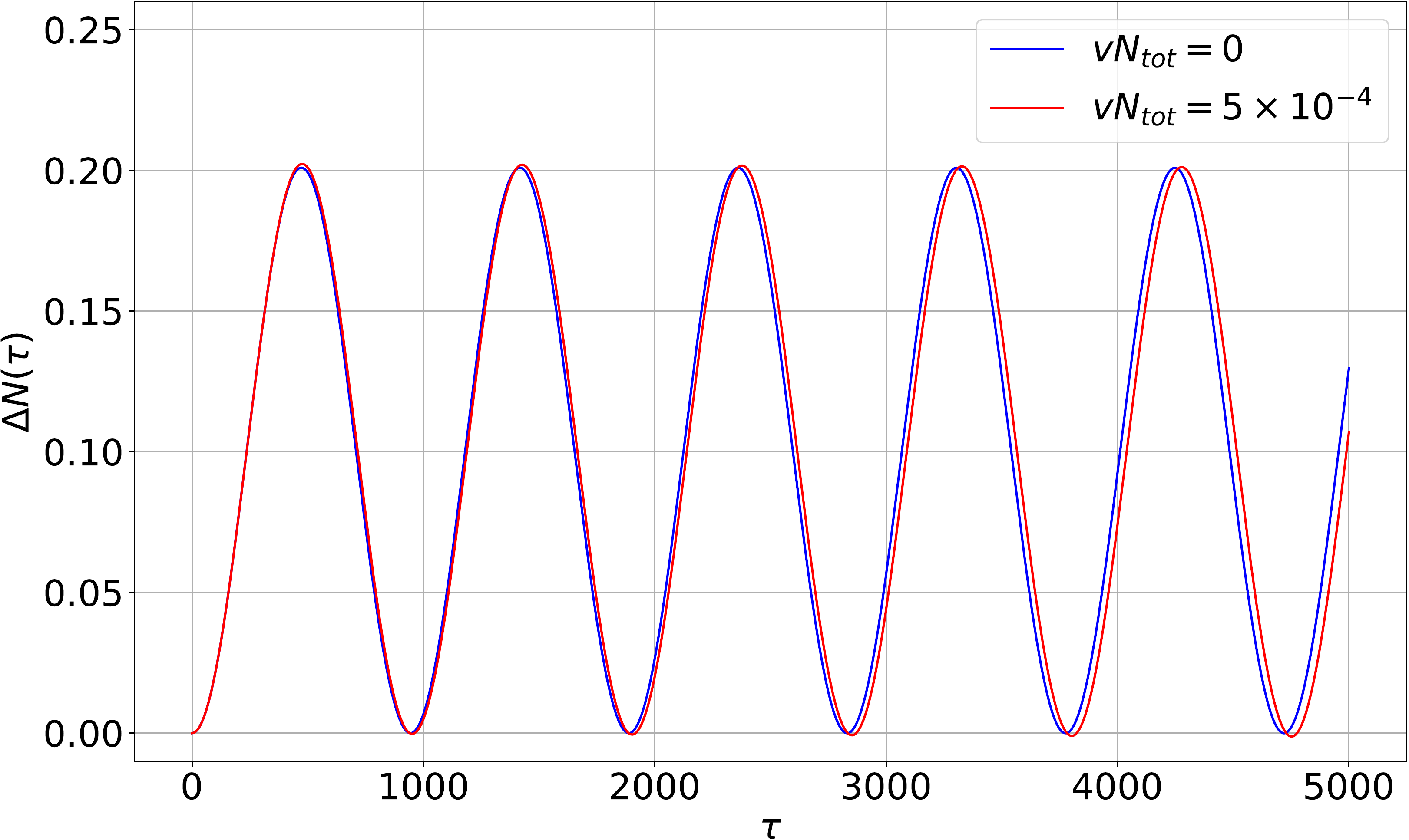}
		}
	 
		\caption{The time evolution of $\Delta N$ with self-interaction or constant $\Delta E$ in low-frequency regime. Parameter setting: $(k,\omega,\theta)=(5\times 10^{-4}, 10^{-2},\arccos(1/3))$.}
		\label{INT_FIG}
	 \end{figure}

	To begin with, we argue that the effect of spin-dependent interaction is similar to that of the spin-independent one. Notice that $\Lambda_{i}$ is always proportional to the $\bm{s}\cdot\bm{n}=+1$ eigenstate along one particular axis $\bm{n}$. Then, one can rotate the axis such that $\bm{e}_z\parallel\bm{n}$. In that case, we have $s_z\Lambda_i=\Lambda_i$ and
	\begin{equation}
		s_{z,i}=\Lambda^{\dg}_is_z\Lambda_i=\Lambda^{\dg}_i\Lambda_i=N_i,\;\:s_{x,i}=s_{y,i}=0,
	\end{equation}
	which further yield
	\begin{equation}
		\begin{aligned}
			v\bm{s}_i\cdot\bm{s}\Lambda_i&=v\Lambda^{\dg}_i\bm{s}\Lambda_i\cdot\bm{s}\Lambda_i\\
			&=vs_{z,i}s_z\Lambda_i\\
			&=vN_i\Lambda_i.
		\end{aligned}
	\end{equation}
	Therefore, the spin-dependent interaction term behaves like the spin-independent one. Intuitively, this is because all the atoms share the same spin wavefunction in one trap and the spin cannot be changed by the mean field of itself. In fact, from numerical results as shown in Fig.\,\ref{INT_u} and Fig.\,\ref{INT_v}, the time evolution of $\Delta N$ in the presence of spin-dependent or spin-independent interactions are identical. Thus it suffices to consider spin-independent interactions. In the following discussions, we exclusively focus on the impact of weak self-interactions. Fortunately, the strength of these self-interactions can be adjusted through Feshbach resonance \cite{feshbach_1,feshbach_observation_1,feshbach_observation_2,optical_feshbach_1}. Consequently, a weak self-interaction, on one hand, can stabilize the superfluid phase \cite{annett_sc_book,landau_criterion}; on the other hand, it will not compromise the qualitative features elucidated in our proposal.
	
	First, we discuss interaction effect in the low-frequency limit. Since we only care about the time evolution of particle numbers, we may consider an equivalent set of equations   
	\begin{equation}
	\begin{cases}
	\;i\pt_{\tau}\Lambda_1=-\bm{s}\cdot\bm{n}(\tau)\Lambda_1-k\Lambda_2+(uN_1-uN_2)\Lambda_1&\\
	\;=(-\bm{s}\cdot\bm{n}(\tau)-uN_{tot}\Delta N)\Lambda_1-k\Lambda_2&\\
	\;i\pt_{\tau}\Lambda_2=-k\Lambda_1-s_z\Lambda_2&
	\end{cases}
	\label{INT_reduced}
	\end{equation}
	which can be done by taking substitution
	\begin{equation}
	\Lambda\rightarrow\exp\left(-i\int^{\tau}u N_2(\lambda)\rmd\lambda\right)\Lambda
	\end{equation}
	Since the interaction strength is small, the $\Delta N$ term in Eq.\,\eqref{INT_reduced} can be substituted by $\Delta N^{(0)}$ (the particle number difference in absence of self-interaction). Thus the interaction effect is equivalent to an asymmetric double-well potential with the difference of ground state energy to be $\Delta E=E_1-E_2=u\Delta N^{(0)}$.
	
	To roughly discuss the effect of a time-dependent energy difference $\Delta E=u\Delta N^{(0)}(\tau)$, let us consider a simplified case: a constant $\Delta E>0$
	\begin{equation}
	\begin{cases}
	\;i\pt_{\tau}\Lambda_1=(-\bm{s}\cdot\bm{n}(\tau)-\Delta E)\Lambda_1-k\Lambda_2&\\
	\;i\pt_{\tau}\Lambda_2=-k\Lambda_1-s_z\Lambda_2&
	\end{cases}
	\label{TOY}
	\end{equation}
	
	In the low-frequency regime, by following the same treatment in Sec.\,\ref{LF}, we get the (approximate) particle number difference
	\begin{equation}
	\begin{aligned}
	&\Delta N\simeq\sin\left(2\arctan\left(\frac{\omega}{\eta k}(1+\Gamma)\right)\right)\\
	&\times\sin^2\left(\frac{1}{2}\omega(1-\cos\theta)\sqrt{1+\left(\frac{\eta k}{\omega}(1-\Gamma)\right)^2}\tau\right)
	\end{aligned}
	\end{equation}
	where
	\begin{equation}
	\Gamma\equiv\frac{\omega(1-\cos\theta)\Delta E}{k^2(1+\cos\theta)^2}>0
	\end{equation}
	Comparing the equation above with Eq.\,\eqref{SBJJ_PREDICTION}, we find that a positive energy difference leads to the increase of period. Consider a system with $k\ll \omega$. In that case, one can neglect the tunneling term in one period of adiabatic rotation and two spinors ($\Lambda_{1/2}$) would keep being parallel to local magnetic fields. Intuitively, after one period, the $\Delta E$ term should lead to a dynamical phase $e^{i\Delta E T}$ on the left side, which decreases the phase difference of the junction and hence inhibits the current. The effect of self-interaction and constant energy difference can be calculated numerically. As is shown in Fig.\,\ref{INT_u} and Fig.\,\ref{INT_dE}, both additions lead to an increase of period, as expected. Clearly, the presence of interaction does not bring about any qualitative change. 

\begin{figure}[tb]
		\subfloat[Spin-independent interaction]
		{  \label{INT_uanharmonic}
		   \includegraphics[width=.42\textwidth]
		   {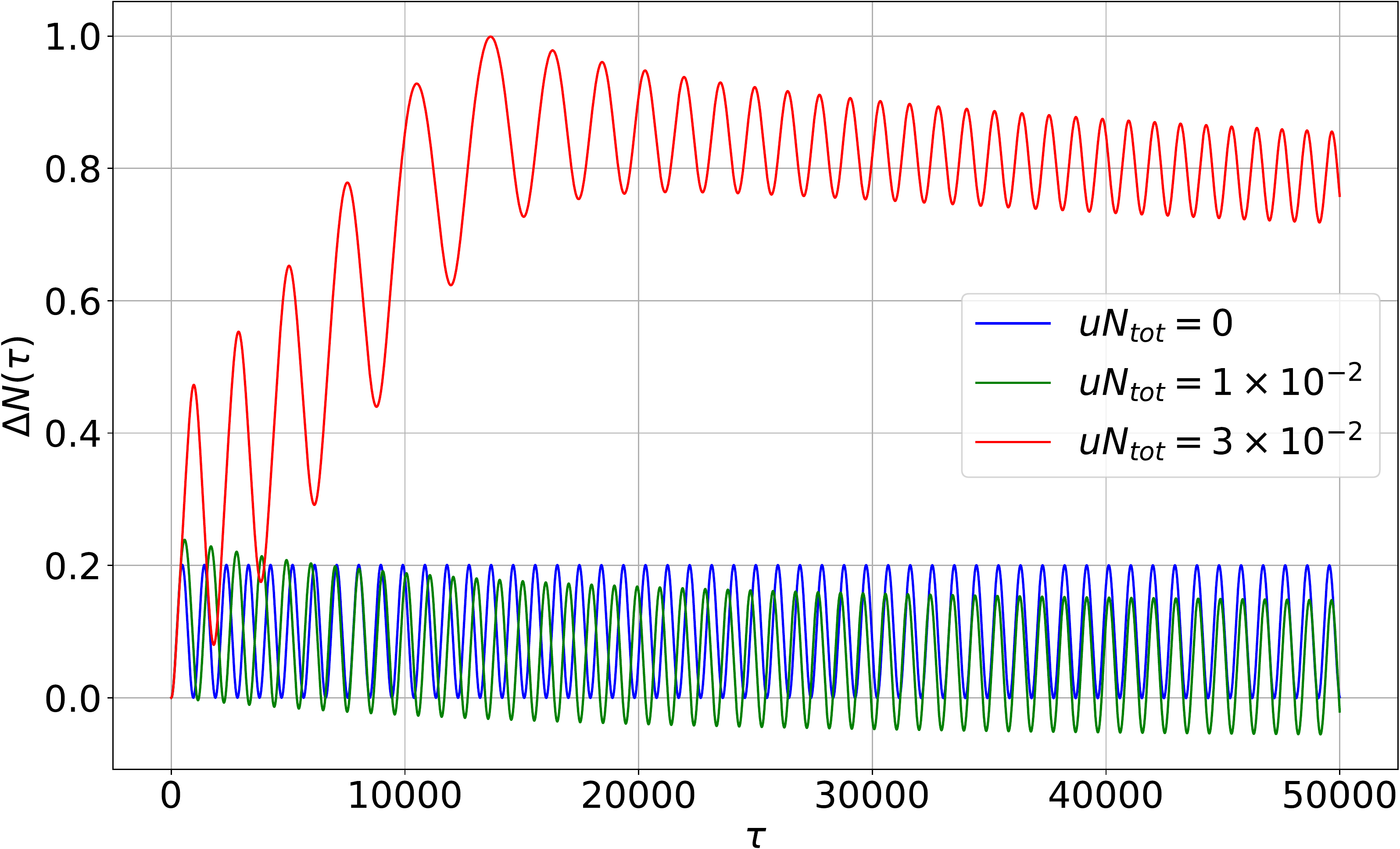}
		}

		\subfloat[Spin-dependent interaction]
		{  \label{INT_vanharmonic}
		   \includegraphics[width=.42\textwidth]
		   {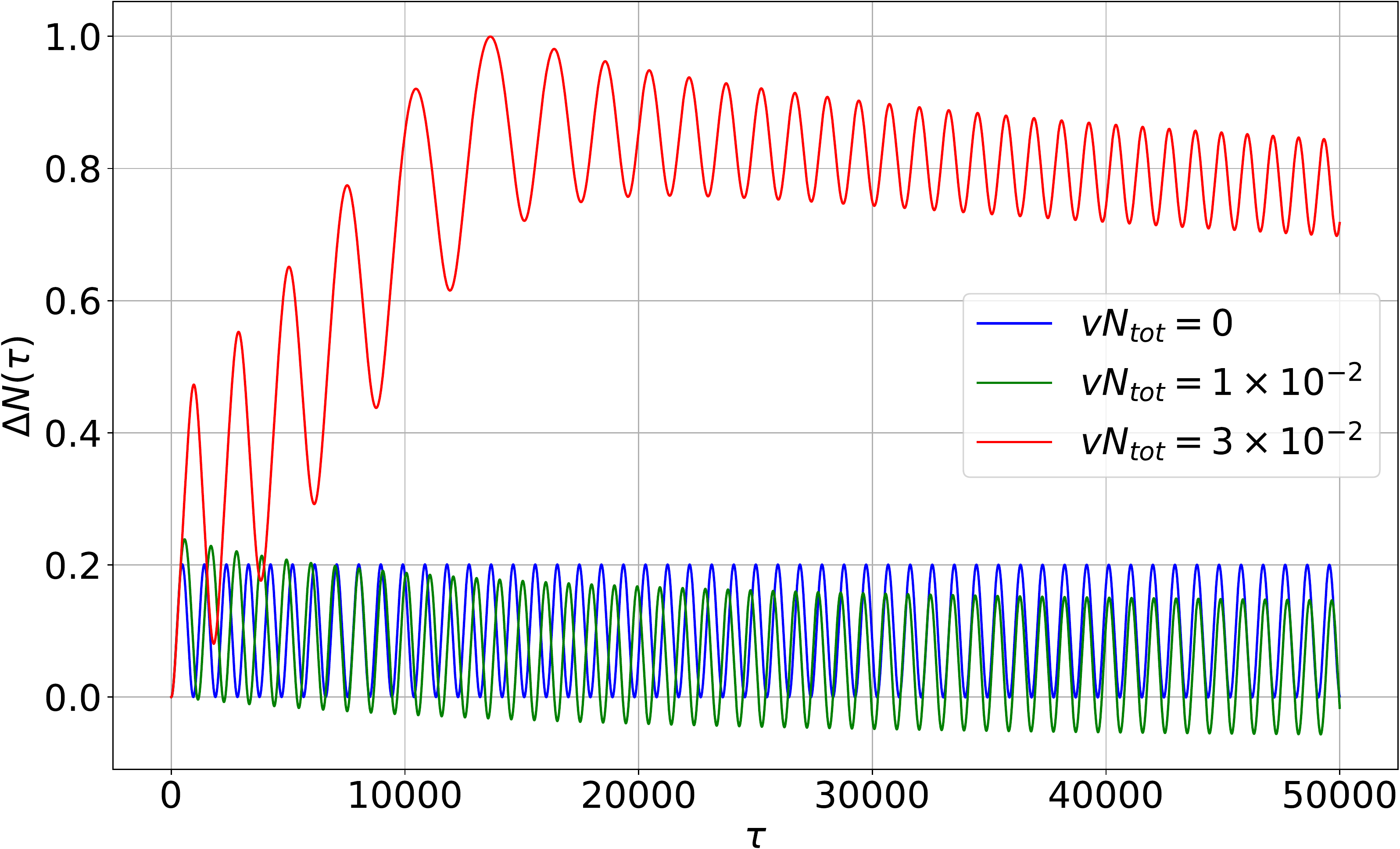}
		}
	 
		\caption{$\Delta N(\tau)$ with larger interaction. Parameter setting: $(k,\omega,\theta)=(5\times 10^{-4}, 10^{-2},\arccos(1/3))$.}
		\label{INT_anharmonics}
	 \end{figure}

	To estimate the strength of the weak interaction, we consider the geometrical and dynamical phases accumulated in one period. The geometrical phase is roughly $\gamma_{g}=2\pi\cos\theta\sim\mathcal{O}(1)$, and the dynamical phase difference is $\Delta\gamma_{d}\sim uN_{tot}\Delta NT\lesssim uN_{tot}T\sim uN_{tot}/\omega$ (and $\Delta\gamma_{d}\lesssim vN_{tot}/\omega$ for spin-dependent interaction). As a result, to obtain a weak interaction, one should require that $\Delta\gamma_{d}\ll \gamma_{g}$, namely $uN_{tot}\ll \omega$. Our argument breaks down if parameters do not satisfy the above condition. In that case, the interaction may not only modify the phase difference, but also significantly change the amplitude of $\Delta N(\tau)$, which could become anharmonic due to the nonlinearity as shown in Fig.\,\ref{INT_uanharmonic} and Fig.\,\ref{INT_vanharmonic}.

	The interaction strength $\alpha$ and $\beta$ can be tuned by the method of Feshbach resonance \cite{feshbach_1,feshbach_2}. Using optically induced Feshbach resonance \cite{optical_feshbach_1,optical_feshbach_2,optical_feshbach_3,optical_feshbach_4}, the scattering length $a_s$ can be modified from $10\:a_0$ to $190\:a_0$ ($a_0$ is the Bohr radius). Noticing that $\alpha$ and $\beta$ have the magnitude of $4\pi\hbar^2a_s/m$ \cite{zhai_2021}, we can estimate the magnitude of interaction energy. Consider a 1D optical trap potential $V(x)=V_0\cos(2kx)$ confining $N_{tot}\sim 10^3$ \ce{^{87}Rb} atoms, with typical value $V_0\sim 300\mathrm{Hz}$ and $\lambda=2\pi/k\sim 10\mathrm{\mu m}$. In that case, the typical length of spatial wavefunction at the bottom of the potential is $\xi\sim 0.7\mathrm{\mu m}$. Hence, the interaction energy is of the magnitude of
	\begin{equation}
		\frac{4\pi\hbar^2a_s}{m}\cdot\frac{N_{tot}}{\xi^3}\gtrsim
		\frac{4\pi\hbar^2\cdot 10a_0}{m}\cdot\frac{N_{tot}}{\xi^3}\sim 10^{-11}\mathrm{eV}
	\end{equation}
	Comparing this with the typical magnitude \cite{zhai_2021} of Zeeman energy ($\sim 10^{-6}\mathrm{eV}$), we find that $uN_{tot}$ and $vN_{tot}$ have the order of $10^{-5}$. Thus, to ensure that the interaction is weak in the low-frequency regime, one should set the rotation frequency of field to be of order $\omega\sim 10^{-3}-10^{-2}$, namely $\Omega\sim 10^5-10^6\mathrm{Hz}$.
	
        In the high-frequency limit, Eq.\,\eqref{INT_reduced} is still valid, and one can still substitute $\Delta N$ by its unperturbed value $\Delta N^{(0)}$. On the other hand, $\Delta N^{(0)}$ oscillates with frequency given by Eq.\,\eqref{bohr_freqs}. Thus it remains almost a constant when taking time average over a period $T\sim 1/\omega$. Therefore, the net contribution of the interaction term in Eq.\,\eqref{INT_reduced} to the Floquet Hamiltonian is a small (time-dependent) energy difference $\delta=-uN_{tot}\Delta N^{(0)}/2$, as shown in Eq.\,\eqref{adding_energy_difference}. Thus, the most significant effect of weak interaction is to modify the oscillation frequency of $\Delta N$ as shown in Eq.\,\eqref{energy_difference_effect}. 
        
\begin{figure}[htb]
		\subfloat[$\Lambda_+=(1/\sqrt{2})(1,1)^T,\;\Lambda_0=\Lambda_-=0$]
		{  \label{HF_INT_1}
		   \includegraphics[width=.45\textwidth]
		   {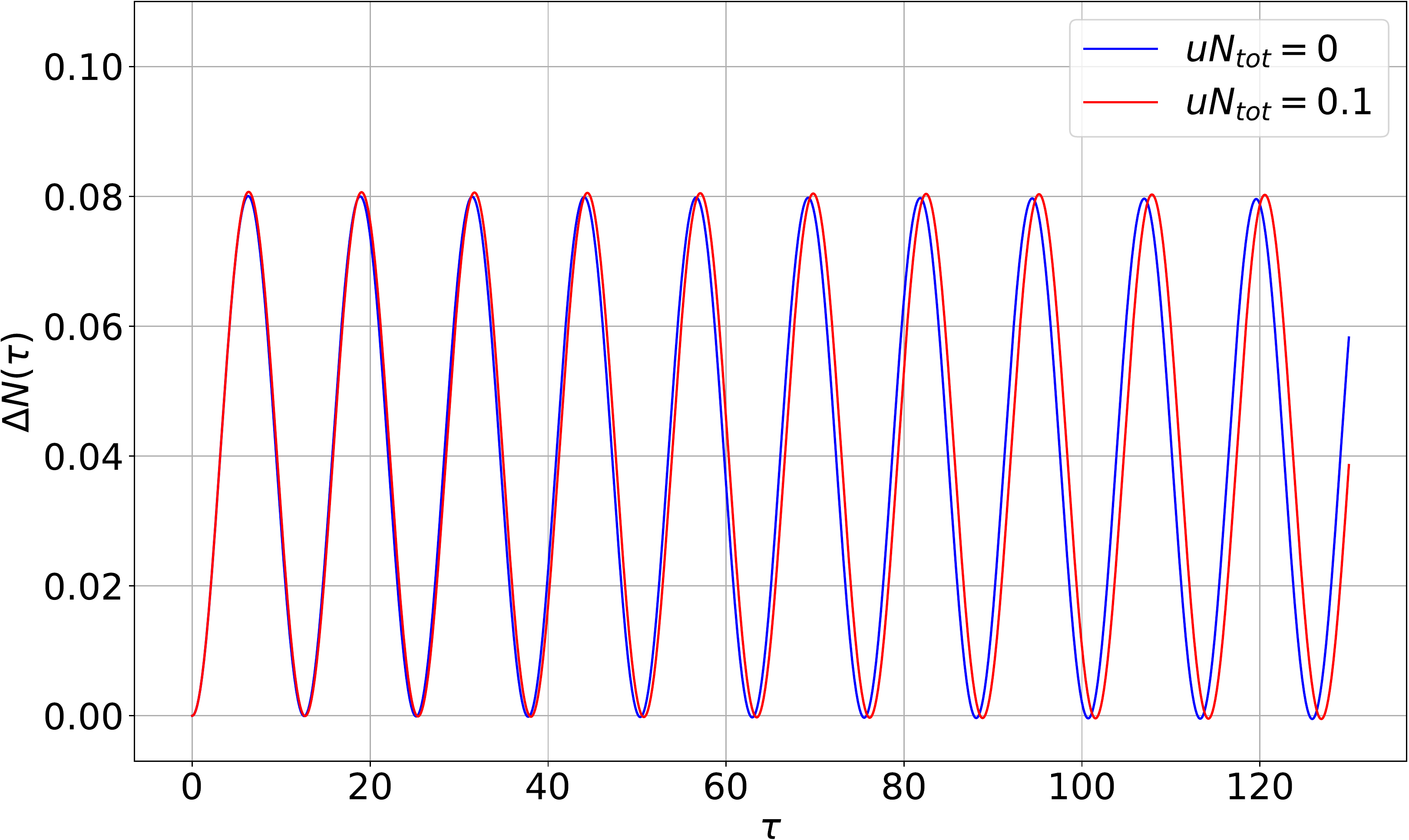}
		}

		\subfloat[$\Lambda_+=(1/\sqrt{2})(1,-1)^T,\;\Lambda_0=\Lambda_-=0$]
		{  \label{HF_INT_2}
		   \includegraphics[width=.45\textwidth]
		   {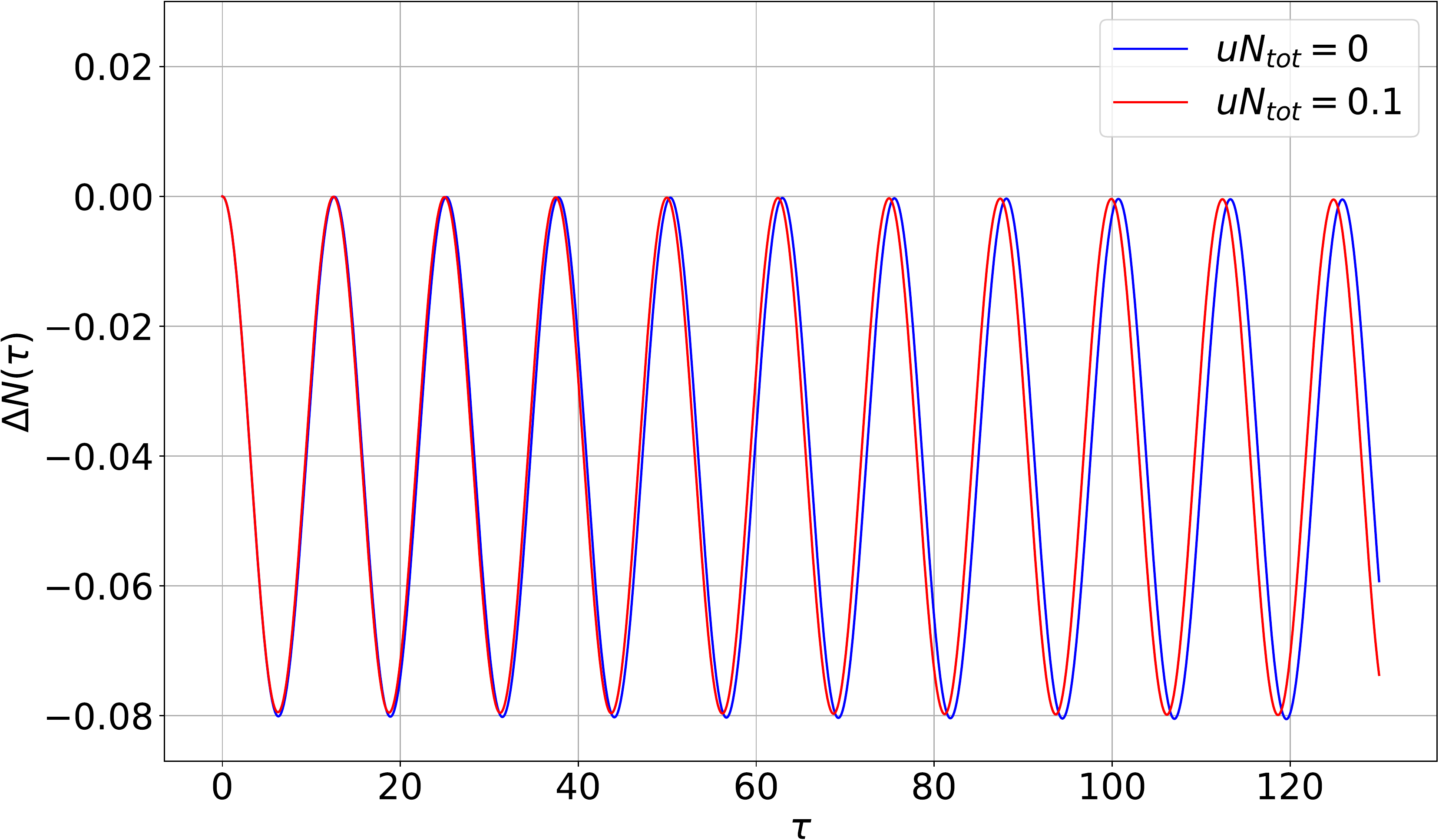}
		}
	 
		\caption{The time evolution of $\Delta N$ with/without self-interaction in high-frequency regime. Parameter setting: $(k,\omega,\theta)=(10^{-2}, 300,\pi/3)$.}
		\label{HF_INTs}
	 \end{figure}

        For oscillations with definite sign of $\Delta N^{(0)}$, we expect that the interaction would have the same qualitative effect as that of Eq.\,\eqref{energy_difference_effect}. Fig.\,\ref{HF_INTs} shows such kind of oscillation of $\Lambda_+$ components with different initial conditions. For Fig.\,\ref{HF_INT_1}, $\Delta N^{(0)}$ is positive, so effectively we have a negative energy difference $\delta$. According to Eq.\,\eqref{energy_difference_effect}, the oscillation frequency ($\omega_+$) decreases as expected. On the contrary, for Fig.\,\ref{HF_INT_2}, $\Delta N^{(0)}$ is negative and $\delta$ is positive, which then increases the oscillation frequency.
        
	\section{Summary}
	\label{Summ}
    We have investigated the particle current in a system composed of spin-1 bosons within a BJJ framework. Our findings suggest that the presence of a rotating magnetic field induces an oscillation in the particle number difference, observable in both low- and high-frequency limits. This oscillation results in a population imbalance, with the potential to achieve maximum bias through the adjustment of specific parameters. In the low-frequency limit, the system's dynamics can be comprehended by applying the adiabatic theorem. Furthermore, by taking a proper limit in that case, the particle current can be intuitively explained as an interplay between the Abelian Berry phase and the Josephson effect.
    In the high-frequency limit, the stroboscopic behavior of the current is encapsulated in the Floquet Hamiltonian, which, in our system, is simply the time average of the original time-dependent Hamiltonian. Additionally, we have also explored the impact of weak self-interactions, which only quantitatively affect the oscillation frequency. We remark that our proposed scheme is intrinsically different from the previous reported  BJJ setup \cite{PhysRevLett.102.185301,PhysRevA.80.023616,internalBJJ_2,PhysRevB.90.144419,PhysRevA.85.043609,PhysRevA.81.063609} and may be straightforwardly generalized for detecting other types of charge neutral condensates, such as the exciton superfluids \cite{osti_7285279,PhysRevLett.93.266801,Jiang2015,PhysRevB.78.041302,PhysRevLett.127.127702}.

	\section*{Acknowledgements}
	The authors are grateful for the previous collaborations with X. C. Xie and Q.-F. Sun. We appreciate support from National Natural Science Foundation of China (NSFC) under Grant No. 23Z031504628, Pujiang Talent Program Grant No. 21PJ1405400, Jiaoda2030 Program Grant No. WH510363001, TDLI starting up grant, and Innovation Program for Quantum Science and Technology Grant No. 2021ZD0301900.
	
	\appendix
	
	\section*{Appendix A: Generalization of the Non-Degenerate Adiabatic Theorem}
	\label{App_A}
	We now present a proof of the generalized adiabatic theorem. As is discussed in Sec.\,\ref{LF}, we consider an "almost" degenerate system with Hamiltonian $\mathcal{H}=\mathcal{H}(\lm(\tau))$ and instantaneous eigenstates are denoted as $\ket{n_{i}(\lm=\lm(\tau))}$. We will continue to use Greek and Latin subscripts to distinguish eigenstates with different zeroth-order energies. However, we use Latin indices starting from $i$  (e.g. $i$, $j$, $k$) to denote a generic instantaneous eigenstate. The state of a system, $\ket{\psi(\tau)}$, satisfies the time-dependent Schr\"{o}dinger equation (with $\hb$ set to 1)
	\begin{equation}
		i\frac{\rmd}{\rmd\tau}\ket{\psi(\tau)}=\mathcal{H}(\lm(\tau))\ket{\psi(\tau)}
		\label{Schro}
	\end{equation}
	
	If one starts with
	\begin{equation}
	\ket{\psi_{\mu}(\tau=0)}\equiv\ket{n_{\mu}(\lm(\tau=0))}
	\end{equation}
	at time $\tau$ we have the expansion
	\begin{equation}
	\begin{aligned}
	&\ket{\psi_{\mu}(\tau)}\\
	&=\sum_{\nu}U_{\mu\nu}(\tau)\ket{n_{\nu}(\lm(\tau))}
	+\sum_{a}U_{\mu a}(\tau)\ket{n_{a}(\lm(\tau))}
	\end{aligned}
	\end{equation}
	with initial condition $U_{\mu j}=\delta_{\mu j}$. Substituting it to Eq.\,\eqref{Schro}, we have
	\begin{equation}
	\begin{aligned}
	&i\sum_{\nu}\left(
	\dot{U}_{\mu\nu}\ket{n_{\nu}}+U_{\mu\nu}\ket{\dot{n}_{\nu}}
	\right)\\
	&+i\sum_{a}\left(
	\dot{U}_{\mu a}\ket{n_{a}}+U_{\mu a}\ket{\dot{n}_{a}}
	\right)\\
	&=\sum_{\nu} U_{\mu\nu}E_{\nu}\ket{n_{\nu}}
	+\sum_{a} U_{\mu a}E_{a}\ket{n_{a}}
	\end{aligned}
	\end{equation}
	We want to show that $U_{\mu a}$'s are suppressed in the adiabatic limit. Taking overlap with $\bra{n_a}$ and rearranging, we have
	\begin{equation}
	\dot{U}_{\mu a}=-iU_{\mu a}E_a-\sum_{\nu} U_{\mu\nu}\inp{n_a}{\dot{n}_{\nu}}
	-\sum_b U_{\mu b}\inp{n_b}{\dot{n}_a}
	\end{equation}
	where we have used the orthogonality of $\ket{n_i}$'s. From initial conditions we know that $U_{\mu a}=0$, so the first and third terms have been suppressed already. We only need to show that terms like 
	\begin{equation}
	U_{\mu\nu}\inp{n_a}{\dot{n}_{\nu}}
	\label{APP_TAR}
	\end{equation}
	are also suppressed. However, this can be done by noticing that
	\begin{equation}
	\begin{aligned}
	\inp{n_a}{\frac{\rmd}{\rmd \tau}(H n_{\mu})}&=\inp{n_a}{\frac{\rmd}{\rmd \tau}(E_{\mu} n_{\mu})}
	=E_{\mu}\inp{n_a}{\dot{n}_{\mu}}\\
	&=\bra{n_a}\dot{H}\ket{n_{\mu}}+\bra{n_a}H\ket{\dot{n}_{\mu}}\\
	&=\bra{n_a}\dot{H}\ket{n_{\mu}}+E_a\inp{n_a}{\dot{n}_{\mu}}\\
	\Rightarrow \inp{n_a}{\dot{n}_{\mu}}&=\frac{1}{E_{\mu}-E_a}\bra{n_a}\dot{H}\ket{n_{\mu}}
	\end{aligned}
	\end{equation}
	Thus, terms in Eq.\,\eqref{APP_TAR} can be suppressed as long as
	\begin{equation}
	\inp{n_a}{\dot{n}_{\nu}}=\frac{1}{E_{\nu}-E_a}\bra{n_a}\dot{H}\ket{n_{\nu}}\ll 1
	\end{equation}
	Since $\ket{n_{\nu}}$ and $\ket{n_a}$ have different zeroth-order energies, one can equally require that the inverse of characteristic time of Hamiltonian is small compared with the zeroth-order energy spacing, which completes the proof.
	
	\section*{Appendix B: Gauge Invariance}
	To complete the argument in Sec.\,\ref{LF}, one still needs to show that the results derived are independent of the set of instantaneous eigenstates we choose. That is, upon a gauge transformation $\ket{n'_a(\phi)}=\Omega_{ab}(\phi)\ket{n_b(\phi)}$ (with $\Omega$ a unitary transformation), the value of $N_1$ and $N_2$ calculated from Eq.\,\eqref{SBJJ_PREDICTION} should be invariant. Strictly speaking, the spectrum of Hamiltonian is non-degenerate, though the gap is rather small. Therefore, the $\Omega$ matrix has to be diagonal:
	\begin{equation}
	\Omega=\begin{pmatrix}
	e^{-i\theta_a} & \\
	& e^{-i\theta_b}
	\end{pmatrix}
	\end{equation}
	where $\theta_a$ and $\theta_b$ are functions of $\phi$ such that $e^{-i\theta_x}|_{\phi=\phi_0+2\pi}=e^{-i\theta_x}|_{\phi=\phi_0}\; (x=a,b)$.
	
	Upon gauge transformations, the non-Abelian Berry connection becomes
	\begin{equation}
	\begin{aligned}
	\mathcal{A}_{\phi}\rightarrow&=\Omega\mathcal{A}_{\phi}\Omega^{\dg}+i\pt_{\phi}\Omega\cdot\Omega^{\dg}\\
	&=\Omega\left(\mathcal{A}_{\phi}+\underbrace{
		\begin{pmatrix}
		\pt_{\phi}\theta_a & \\
		& \pt_{\phi}\theta_b
		\end{pmatrix}
	}_{\equiv S}
	\right)\Omega^{\dg}
	\end{aligned}
	\end{equation}
	where we have introduced $S$ just for conciseness. Accordingly, the differential equation for the $U$ matrix [Eq.\,\eqref{SBJJ_EOMU_1}] becomes
	\begin{equation}
	\begin{aligned}
	(U^{\dg})'&=-i(\underbrace{\mathcal{A}_{\phi}-\frac{|z|}{\omega}\sg_3}_{\equiv Q})U^{\dg}\\
	\rightarrow (\tilde{U}^{\dg})'&=-i\left[
	\Omega\left(\mathcal{A}_{\phi}+
	S
	\right)\Omega^{\dg}
	-\frac{|z|}{\omega}\sg_3
	\right]\tilde{U}^{\dg}\\
	&=-i\left(\Omega(Q+S)\Omega^{\dg}\right)\tilde{U}^{\dg}
	\end{aligned}
	\label{SBJJ_Gauge_deri_1}
	\end{equation}
	with the initial condition $\tilde{U}^{\dg}(\phi=0)=\tilde{U}(\phi=0)=1$. Note that $\Omega$ and $S$ are functions of $\phi$ rather than constants.
	
	In fact, with $\Omega'=-iS\Omega=-i\Omega S$, which can be easily checked, one can verify that
	\begin{equation}
	\begin{aligned}
	&\left(\Omega(\phi)U^{\dg}(\phi)\Omega^{\dg}(0)\right)'\\
	&=-i\Omega(\phi)S(\phi)U^{\dg}(\phi)\Omega^{\dg}(0)+\Omega(\phi)(-iQU^{\dg}(\phi))\Omega^{\dg}(0)\\
	&=-i\Omega(\phi)(Q+S(\phi))U^{\dg}(\phi)\Omega^{\dg}(0)\\
	&=-i\left(\Omega(\phi)(Q+S(\phi))\Omega^{\dg}(\phi)\right)\cdot\left(\Omega(\phi)U^{\dg}(\phi)\Omega^{\dg}(0)\right)
	\end{aligned}
	\end{equation}	
	which implies that
	\begin{equation}
	\begin{aligned}
	&U^{\dg}(\phi)\rightarrow \tilde{U}^{\dg}(\phi)=\Omega(\phi)U^{\dg}(\phi)\Omega^{\dg}(0)\\
	\Leftrightarrow\; 
	&U(\phi)\rightarrow\tilde{U}(\phi)=\Omega(0)U(\phi)\Omega^{\dg}(\phi)
	\end{aligned}	
	\end{equation}
	by the uniqueness of solution to the differential equation \eqref{SBJJ_Gauge_deri_1}.
	
	On the other hand, the coefficients for the linear combination of state kets also change under the gauge transformation. In fact, noticing that
	\begin{equation}
	\begin{aligned}
	\ket{\psi(0)}&=c_a(0)\ket{n_a(\phi(\tau=0))}+c_b(0)\ket{n_b(\phi(\tau=0))}\\
	&=c_a(0)e^{i\theta_a(0)}e^{-i\theta_a(0)}\ket{n_a(\phi(\tau=0))}\\
	&+c_b(0)e^{i\theta_b(0)}e^{-i\theta_b(0)}\ket{n_b(\phi(\tau=0))}
	\end{aligned}
	\end{equation}
	implies that
	\begin{equation}
	\begin{aligned}
	&(c_a(0),c_b(0))\\
	&\rightarrow (c_a(0)e^{i\theta_a(0)},c_b(0)e^{i\theta_b(0)})=(c_a(0),c_b(0))\Omega^{\dg}(0)
	\end{aligned}
	\end{equation}
	Thus, after evolution, these coefficients become
	\begin{equation}
	\begin{aligned}
	(\tilde{c}_a(\tau),\tilde{c}_b(\tau))
	&=(c_a(0),c_b(0))\Omega^{\dg}(0)\cdot\Omega(0) U(\phi)\Omega^{\dg}(\phi)\\
	&=(c_a(0),c_b(0))U(\phi)\Omega^{\dg}(\phi)\\
	&=(c_a(\tau),c_b(\tau))\Omega^{\dg}(\phi)\\
	&=(c_a(\tau)e^{i\theta_a},c_b(\tau)e^{i\theta_b})
	\end{aligned}
	\end{equation}
	
	Therefore, the state ket (at time $\tau$) calculated through our method
	\begin{equation}
	\begin{aligned}
	\ket{\tilde{\psi}(\tau)}
	&=\tilde{c}_a(\tau)e^{-i\theta_a}\ket{n_a(\phi)}+
	\tilde{c}_b(\tau)e^{-i\theta_b}\ket{n_b(\phi)}\\
	&=c_a(\tau)\ket{n_a(\phi)}+c_b(\tau)\ket{n_b(\phi)}=\ket{\psi(\tau)}
	\end{aligned}
	\end{equation}
	is a gauge invariant, which also implies that quantities which can be calculated directly through $\ket{\psi(\tau)}$ (such as $N_1,\; N_2,\;\mbox{and}\;\Delta N$) are all gauge invariants. 
	
	\bibliographystyle{elsarticle-num}
	\bibliography{reference}
\end{document}